\documentclass[manuscript,screen]{acmart}

\usepackage{booktabs}
\usepackage{listings}
\usepackage{xcolor}
\usepackage[ruled,vlined,linesnumbered]{algorithm2e}
\usepackage{tikz}
\usepackage{cleveref}
\usepackage{mathpartir}
\usepackage{soul}
\usepackage{xurl}
\usepackage{enumitem}
\usepackage{easy-todo}
\usetikzlibrary{shapes,arrows,positioning,fit,backgrounds}

\newcommand{\approach}{\textsf{MifDroid}\xspace}
\newcommand{\etal}{\textit{et al.}\xspace}

\providecommand{\TSA}{\mathcal{A}}
\providecommand{\TSEff}[2]{\langle #1,#2\rangle}
\providecommand{\TSpc}{\mathit{pc}}
\providecommand{\TSpct}{\mathit{pc}_t}
\providecommand{\TSpctf}{\mathit{pc}_{t,f}}
\providecommand{\TSJudge}[5][\TSpc;\TSpct]{#1;#2;#3 \vdash #4 : #5}
\providecommand{\TSAlarm}[2]{\mathsf{alarms}(#1,#2)}
\providecommand{\TSInt}{\mathcal{I}}
\providecommand{\TSPar}{\mathit{Par}}
\providecommand{\TSJoinT}{\cup}
\providecommand{\TSLeqT}{\subseteq}

\providecommand{\TSMArrow}[5]{(#1) \xrightarrow{#2,\;\TSEff{#3}{#4}} #5}
\providecommand{\TSA}{\mathcal{A}}

\providecommand{\wkind}{\mathrm{kind}}

\lstdefinelanguage{JavaScript}{
  keywords={typeof, new, true, false, catch, function, return, null, catch, switch, var, if, in, while, do, else, case, break},
  keywordstyle=\color{blue}\bfseries,
  ndkeywords={class, export, boolean, throw, implements, import, this},
  ndkeywordstyle=\color{darkgray}\bfseries,
  identifierstyle=\color{black},
  sensitive=false,
  comment=[l]{//},
  morecomment=[s]{/*}{*/},
  commentstyle=\color{gray}\ttfamily,
  stringstyle=\color{red}\ttfamily,
  morestring=[b]',
  morestring=[b]"
}

\lstdefinestyle{javastyle}{
  language=Java,
  basicstyle=\ttfamily\small,
  keywordstyle=\color{blue}\bfseries,
  commentstyle=\color{gray}\itshape,
  stringstyle=\color{red},
  numbers=none,
  backgroundcolor=\color{gray!5},
  showspaces=false,
  showstringspaces=false,
  showtabs=false,
  frame=single,
  rulecolor=\color{black!30},
  tabsize=2,
  captionpos=b,
  breaklines=true,
  breakatwhitespace=true,
  escapeinside={(*@}{@*)},
  xleftmargin=5pt,
  xrightmargin=5pt,
  aboveskip=8pt,
  belowskip=8pt,
  columns=flexible
}

\lstdefinestyle{jsstyle}{
  language=JavaScript,
  basicstyle=\ttfamily\small,
  keywordstyle=\color{blue}\bfseries,
  commentstyle=\color{gray}\itshape,
  stringstyle=\color{red},
  numbers=none,
  backgroundcolor=\color{gray!5},
  showspaces=false,
  showstringspaces=false,
  showtabs=false,
  frame=single,
  rulecolor=\color{black!30},
  tabsize=2,
  captionpos=b,
  breaklines=true,
  breakatwhitespace=true,
  escapeinside={(*@}{@*)},
  xleftmargin=5pt,
  xrightmargin=5pt,
  aboveskip=8pt,
  belowskip=8pt,
  columns=flexible
}

\usepackage[most]{tcolorbox}
\usepackage{xcolor}

\definecolor{asegraybg}{RGB}{248,248,248}
\definecolor{asegrayborder}{RGB}{170,170,170}
\newtcolorbox{findingbox}{
	enhanced,
	breakable,
	colback=gray!6,
	colframe=gray!50,
	boxrule=0.4pt,
	arc=1pt,
	left=4pt,
	right=4pt,
	top=1pt,
	bottom=1pt
}
\acmJournal{TOSEM}

\begin{document}
\sloppy
\title{A Hybrid Information Flow Analysis Framework for Android Hybrid Apps}

\author{Jyoti Prakash}
\affiliation{%
  \institution{University of Southern Denmark}
  \city{Odense}
  \country{Denmark}
}
\email{jyp@mmmi.sdu.dk}
\author{Mikkel Baun Kjærgaard}
\affiliation{%
	\institution{University of Southern Denmark}
	\city{Odense}
	\country{Denmark}
}
\email{mbkj@mmmi.sdu.dk}
\author{Abhishek Tiwari}
\affiliation{%
	\institution{University of Southern Denmark}
	\city{Odense}
	\country{Denmark}
}
\email{abti@mmmi.sdu.dk}

\begin{abstract}
The \texttt{WebView} interface in Android (hybrid) applications enables the native side (e.g., Java) to exchange data with the Web (e.g., JavaScript). This creates cross-language information flow channels that existing single-language analyzers cannot analyze. Existing hybrid app analyses address only part of the problem: some build bridge-aware call graphs but provide limited information-flow analysis, some rely on coarse assumptions, and some focus only on web vulnerabilities. We present \approach{}, a static-analysis-guided dynamic-analysis framework with a lifecycle-aware per-WebView call graph, a bridge-aware type-and-effect system, and a modular JavaScript analysis engine. We evaluated MifDroid on 338 apps from the F-Droid repository (out of 4,126 total) that use \texttt{addJavascriptInterface}, enabling active communication between Java and JavaScript. Our guided dynamic phase adds 662 entries to the 444 produced by static analysis, raising the resolution rate from 36.5\% to 74.5\%, produces the resolved bridge-related entries for 115 apps that static analysis leaves empty. \approach{} reported 14 vulnerable cross-language information flows, five of which have been accepted and fixed by the maintainers.
\end{abstract}

\ccsdesc[500]{Security and privacy~Information flow control}
\ccsdesc[500]{Theory of computation~Program analysis}
\ccsdesc[300]{Software and its engineering~Dynamic analysis}

\keywords{information flow, static and dynamic analysis, Android WebView}

\maketitle
\section{Introduction}

Android applications (apps) are increasingly hybrid, with Java/Kotlin code on the native side (as the host) and JavaScript on the Web (as the guest) communicating through a shared bridge. This communication is facilitated by the \texttt{WebView}~\cite{webview}. The host side can invoke JavaScript/Web via dedicated APIs, such as \texttt{loadUrl} or \texttt{evaluateJavascript}. The communication is not merely one-sided. WebView also enables the guest side (JavaScript) to access the native side functionality via a special host object. We refer to this unique host object as \emph{bridge object}. Although these bridge objects belong to the native-side Java/Kotlin class, they are exposed to JavaScript as if they were its own objects. Bridge objects were introduced to facilitate code reuse across the native side and the web. Unfortunately, this host-to-guest bridge communication also exposes sensitive host data (e.g., location) to a potentially untrusted web page~\cite{ludroid-journal, ifc-analysis-issre}.

Detecting such flows is difficult for two reasons. First, a precise call graph for a hybrid app depends on communication endpoints of the WebView, such as loaded URLs and injected scripts. In practice, these values are often computed at runtime~\cite{omegatest}, for example, through string concatenation, resource lookup, or framework code. Consequently, static analyses may miss bridge communication entirely or merge multiple WebView instances to over-approximate communication context~\cite{ifc-analysis-issre}. Second, dynamic analysis suffers from coverage issues. Even strong test-generation techniques, such as $\omega$Test~\cite{omegatest}, cannot guarantee that all WebView-relevant paths will be exercised. A purely dynamic approach may therefore miss a flow simply because the triggering execution never occurs.

Existing hybrid app analyses only address part of this problem. \textsf{HybriDroid}~\cite{hybdridroid} presented semantics of Java-JavaScript communication. However, their analysis mainly focused on type-mismatch bugs. \textsf{LUDroid}~\cite{ludroid-journal} targeted information flows via the Java-JavaScript bridge. It treats all sensitive data that leaves the host as a leak, flooding results with false positives. The most relevant related work~\cite{ifc-analysis-issre} proposes a demand-driven IFC analysis for WebView bridges. However, their approach is based on static analysis, leaving dynamically computed targets unresolved.  Additionally, their analysis relies on pre-existing call graphs that do not model WebView-specific lifecycle operations.  \textsf{WebViewJSdetect}~\cite{webViewJsDetect} advances the state of the art on the JavaScript side by leveraging concurrent abstract interpretation to detect vulnerabilities, such as unauthorized API access and privilege escalation, within the WebView's execution context. However, it operates entirely within the guest and does not precisely model flows originated on the host. 

In this paper, we present \approach, a static-analysis-guided dynamic-analysis framework for detecting cross-language information flows in Android hybrid apps. The main insight is that static and dynamic analysis are not alternatives but collaborators. The static phase identifies which components host WebViews, collects statically resolvable information, and marks the statically unresolved sites. The dynamic phase is then guided toward those sites to obtain concrete runtime values.  This resolved data is leveraged to create a lifecycle-aware, per-WebView call graph. A type-and-effect system leverages this  call graph to compute confidentiality and integrity effects at the bridge boundary. Finally, a modular JavaScript analysis phase translates the computed effects into synthetic stubs so that existing JavaScript analyzers can detect information flows.

In summary, this paper makes the following contributions:

\smallskip\noindent\textbf{Contributions.}
\begin{enumerate}
	\item A \emph{hybrid analysis} that constructs a per-WebView summary by combining static analysis, guided dynamic analysis, and conservative approximation.
	\item A \emph{lifecycle-aware, per-WebView call graph} that builds one communication context for each WebView instance. This avoids spurious cross-WebView edges and models WebView-specific communication semantics.
	\item A \emph{type-and-effect system} that runs on the program closed by the static and dynamic resolution of the first phase. It summarizes each bridge method by a confidentiality effect, recording whether the method may return a confidential value, and by integrity alarms, recording whether guest-controlled arguments may reach internal host state. This enables static detection of cross-language information flow violations.
	\item A \emph{modular JavaScript analysis} that translates effect summaries into instrumented stubs. This reduces cross-language flows to a standard single-language taint analysis.
	\item An evaluation on 338 large-scale F-Droid apps that exercise the WebView bridge. We show that guided dynamic analysis adds 662 entries to the 444 produced by static analysis, raising the resolution rate from 36.5\% to 74.5\%, finds many \texttt{evaluateJavascript} targets that static analysis misses, resolves bridge-related entries for 115 apps that static analysis leaves empty. \approach{} reported 14 vulnerable cross-language information flows. At the time of submission, five have been accepted and fixed by the corresponding maintainers, and the remaining are still open.
\end{enumerate}

\section{Motivation}\label{sec:motivation}

Information-flow analysis of an Android hybrid app must follow data through both native code and JavaScript~\cite{hybdridroid,ifc-analysis-issre}. A bridge method may return sensitive native data to JavaScript or use a JavaScript argument to update native state. Tracking these flows requires determining which scripts can call which native methods through each WebView.

Listing~\ref{lst:running} illustrates this problem with a simplified credential leak. The Activity reads the interface name from \texttt{R.string.bridge\_name} (line~\ref{ln:running:bridge-name}) and registers an \texttt{API} object under that name (line~\ref{l1}). It constructs the URL and loads a bundled HTML file on lines~\ref{ln:running:url-prefix}--\ref{ln:running:load}. Assume that the resource contains \texttt{"bridge"} and that \texttt{Config.PAGE} depends on runtime state, with value \texttt{"app"} in the execution shown. We label \texttt{password} on line~\ref{ln:running:password} as sensitive ($H$) and treat \texttt{System.out.println} on line~\ref{ln:running:sink} as a public sink ($L$). In \texttt{app.html}, \texttt{bridge.getPassword()} (line~\ref{ln:running:js-get-password}) calls the native method on line~\ref{ln:running:get-password} and returns the password to JavaScript. JavaScript then passes it back to Java through \texttt{bridge.log(pwd)} (line~\ref{ln:running:js-log}), which reaches the public sink. The flow crosses the language boundary twice before reaching the sink.

\begin{lstlisting}[caption={Java (top) and JavaScript (bottom).},label=lst:running, float=b, numbers=left, escapechar=|]
class API {
  private String password = "secret123"; |\label{ln:running:password}\textcolor{gray}{// H}|
  @JavascriptInterface
  public String getPassword() { return password; } |\label{ln:running:get-password}|
  @JavascriptInterface
  public void log(String msg) {
    System.out.println(msg); |\label{ln:running:sink}\textcolor{gray}{// L sink}|
  }
}
class MainActivity extends Activity {
  protected void onCreate(Bundle state) {
    WebView wv = new WebView(this); |\label{ln:running:webview-allocation}|
    wv.getSettings().setJavaScriptEnabled(true);
    String prefix = getString(R.string.bridge_name); |\label{ln:running:bridge-name}|
    wv.addJavascriptInterface(new API(), prefix); | \label{l1}|
    String url = "file:///android_asset/" + |\label{ln:running:url-prefix}|
        Config.PAGE + ".html"; | \label{l2}|
    wv.loadUrl(url); |\label{ln:running:load}|
  }
}
// --- app.html ---
var pwd = bridge.getPassword(); |\label{ln:running:js-get-password}\textcolor{gray}{// pwd: H}|
bridge.log(pwd);                |\label{ln:running:js-log}\textcolor{gray}{// H $\rightarrow$ L}|
\end{lstlisting}

A call graph is crucial for tracking this flow. The target of \texttt{bridge.getPassword()} depends on the object and name passed to \texttt{addJavascriptInterface}, together with the code loaded into the WebView. Static analysis can recover \texttt{bridge} from the resource table, but following \texttt{Config.PAGE} may yield only the URL prefix and suffix. Helper methods and runtime inputs can leave interface names or injected scripts unresolved~\cite{ludroid-journal}. Omitting an unresolved interaction can miss the flow; connecting it to every bridge in the app can introduce infeasible paths.

The WebView instance also matters. In Listing~\ref{lst:running}, interface registration on line~\ref{l1} and page loading on line~\ref{ln:running:load} use the same WebView, \texttt{wv}. Suppose a second WebView loads another page and registers a different object under the same name \texttt{bridge}. Calls through \texttt{bridge} in that page refer to the second object, not the \texttt{API} object registered on \texttt{wv}. An app-wide bridge model can incorrectly connect those calls to \texttt{API.getPassword()} in the first WebView. The analysis therefore needs to associate registrations and loaded scripts with the corresponding WebView instances. It must also track aliases, since two variables can refer to the same WebView instance.

Execution order adds another difficulty. Tiwari \etal show that updates to a registered object's fields after \texttt{loadUrl} can affect the JavaScript loaded by that call~\cite{ifc-analysis-issre}. Consider a variant of Listing~\ref{lst:running} where the \texttt{API} class has a native setter for \texttt{password} and the Activity keeps a reference to the registered object. Immediately after \texttt{loadUrl} on line~\ref{ln:running:load}, the Activity uses the setter to store a new sensitive password. The JavaScript call on line~\ref{ln:running:js-get-password} may receive this password through the getter on line~\ref{ln:running:get-password}, even though the native write appears after the loading call. An analysis that considers only the object's state at registration or loading can miss the dependency from this later write to the getter's return value. The analysis must therefore include subsequent native writes that may affect exposed bridge methods.

Dynamic execution can record concrete interface names, URLs, and injected scripts, but only for the paths it executes. Reaching the relevant WebView can require a sequence of UI actions or input supplied in another Activity~\cite{omegatest}. In Listing~\ref{lst:running}, an execution with \texttt{Config.PAGE="app"} records \nolinkurl{file:///android_asset/app.html} at the \texttt{loadUrl} call on line~\ref{ln:running:load}. This observation does not establish which other pages the same call may load in different executions.

The observed values must also be matched to native call sites. Suppose the same Activity also contains \texttt{wv2.loadUrl(otherUrl)}, which loads only \texttt{help.html} into a second WebView, \texttt{wv2}. Assume static analysis cannot determine either filename. A log that records only a page URL and \texttt{MainActivity} does not identify which of the two calls produced that entry. If the observed URL of \texttt{app.html} is retained for both calls, the analysis treats that page as possibly loaded into \texttt{wv2} as well. Suppose \texttt{wv2} registers a different object as \texttt{bridge}, with its own \texttt{getPassword} method. The JavaScript call on line~\ref{ln:running:js-get-password} can then be incorrectly connected to that object's getter, even though \texttt{app.html} never runs in \texttt{wv2}. Dynamic refinement must therefore account for both execution coverage and ambiguity when matching observations to static call sites.

\approach\ uses static analysis to recover interface names, URLs, and JavaScript code passed to WebView APIs. It records arguments that remain unresolved and guides execution of the instrumented app toward the Activities and Fragments that create the relevant WebViews. The observed values are matched to unresolved call sites using the API, the Activity or Fragment, and any string fragments recovered statically. If several sites match, each keeps the observed value as a candidate. For each WebView allocation site, the analysis constructs a summary of the bound native objects, their interface names, and the JavaScript code that may execute there. The native analysis also identifies the exposed bridge methods and field updates that may affect those methods. Together, this information is used to connect JavaScript calls to native bridge methods in the inter-language call graph. Arguments that remain unresolved are handled by the fallback rules described in the next section.

\section{Approach}\label{sec:approach}

\begin{figure*}[tbp]
	\centering
	\includegraphics[width=\linewidth]{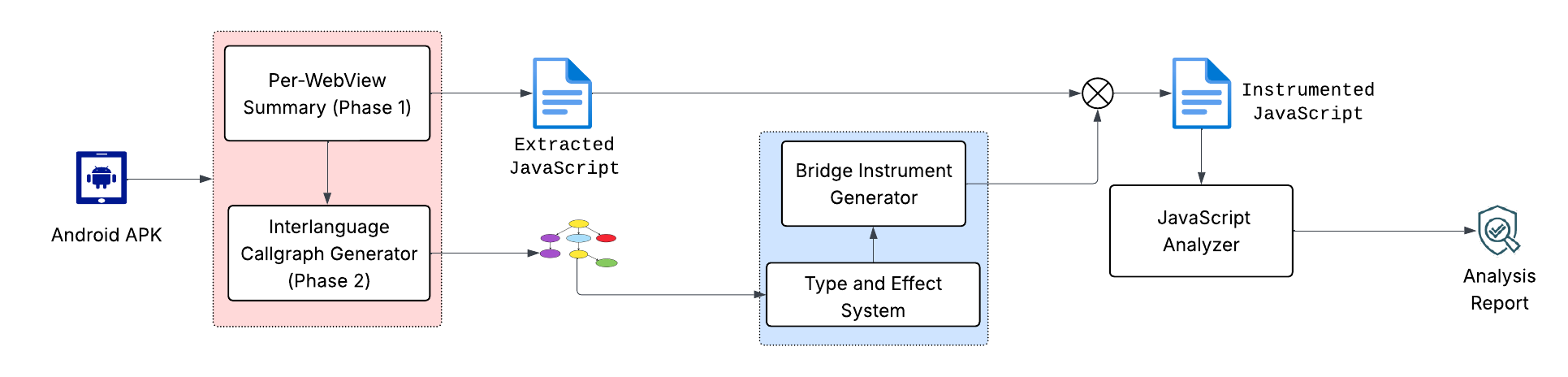}
	\caption{Overall Framework of~\approach}
	\label{fig:framework}
\end{figure*}

Figure~\ref{fig:framework} gives an overview of~\approach. Our analysis is centered on the \emph{WebView instance}, since this is the point where Java--JavaScript communication is configured in a hybrid app. A WebView instance determines which interface objects are shared with JavaScript, under which names they are exposed, which bridge methods and object-state updates become visible through them, and which JavaScript artifacts may execute in the same communication context. Recovering this information is challenging because interface names, loaded URLs, injected scripts, and object updates are often computed at runtime. In addition, WebView follows peculiar semantics, such as asynchronous interaction and flow-insensitive updates of interface objects~\cite{ifc-analysis-issre}.

Our approach has four steps. First, we construct a \emph{per-WebView summary} (detailed in \Cref{sec:stringres}). The summary captures the communication relation of a WebView instance: the interface summaries attached to it, where each such summary records the interface name, the bound object, its exposed bridge methods, and the relevant updates on that object, together with the JavaScript artifacts that may run in the same WebView. Second, we use these summaries to build a \emph{lifecycle-aware inter-language call graph} (\Cref{sec:callgraph}). Third, we run a type-and-effect analysis to compute the type, sensitivity, and effects of the bridge methods. Finally, we project these effects to JavaScript and reuse existing single-language analyzers to check JavaScript-side safety (\Cref{sec:modular-js}).

\subsection{Phase 1: WebView Summary Construction}\label{sec:stringres}

In this phase, we construct a summary for each WebView instance. The summary captures the part of the native-side code that is relevant for Java--JavaScript communication. It records which interface objects may be attached to the WebView, under which names they are visible to JavaScript, which bridge methods they expose, which updates on these objects may affect communication, and which JavaScript artifacts may execute in the same WebView. At this stage, we do not need the full behavior of the app. We only need the communication-relevant part of the WebView and its associated native-side state.

We begin by identifying the WebView API calls that establish this communication. These include \texttt{addJavascriptInterface} for bridge registration, \texttt{loadUrl}, \texttt{loadData}, \texttt{loadDataWithBaseURL}, and \texttt{evaluateJavascript} for loading web content, including JavaScript. For each WebView instance, we keep track of the calls that may act on it and trace them backward in the native-side code. This recovers interface names, interface-object bindings, updates on bound interface objects, loaded resources, and injected scripts. Since these values are often resolved at runtime, Phase~1 proceeds in three steps: static value recovery, guided dynamic refinement, and conservative approximation.

\begin{definition}[Tracked WebView APIs]\label{def:apis}
	The native side configures Java--JavaScript communication through the WebView APIs below. Each row fixes the kind $\wkind(c)$ of a call and the string argument $\mathrm{arg}(c)$ that we record for it. The kind is determined by the invoked API alone.  For \texttt{addJavascriptInterface}, we also record the bound native-side object as $\mathrm{bridgedObj}(c)$, since the interface name alone is not enough to recover the exposed bridge methods and the updates on that object.
	\begin{center}\small
	\begin{tabular}{@{}lll@{}}
		\toprule
		$\wkind(c)$ & WebView API $\mathrm{wvApi}(c)$ & Recorded argument $\mathrm{arg}(c)$ \\
		\midrule
		$\mathsf{bridge}$ & \texttt{addJavascriptInterface(o, n)} & $n$, the interface name \\
		$\mathsf{load}$   & \texttt{loadUrl(u)}                   & $u$, the loaded URL \\
		                  & \texttt{loadData(d, \ldots)}          & $d$, the inline content \\
		                  & \texttt{loadDataWithBaseURL(b, d, \ldots)} & $b$, the base URL, \emph{and} $d$, the inline content \\
		$\mathsf{inject}$ & \texttt{evaluateJavascript(s, \ldots)} & $s$, the injected JavaScript code \\
		\bottomrule
	\end{tabular}
	\end{center}

\end{definition}

\begin{definition}[Communication Sites]\label{def:sites}
	A \emph{communication site} of an app~$a$ is a pair $c = (\ell, x)$ of a WebView API call~$\ell$ occurring in~$a$ and one of the arguments~$x$ that \Cref{def:apis} records at~$\ell$; a call contributes one site per recorded argument, and $C$ collects them all. We write $\mathrm{arg}(c)$ for the recorded argument~$x$, $\mathrm{wvRecv}(c)$ for the WebView reference that~$\ell$ is invoked on, and $\mathrm{encl}(c)$ for the native-side method whose body contains~$\ell$. For the call \texttt{wv.loadUrl(url)} of Listing~\ref{lst:running},
	\[
	  \mathrm{arg}(c) = \texttt{url}, \qquad
	  \mathrm{wvRecv}(c) = \texttt{wv}, \qquad
	  \mathrm{encl}(c) = \texttt{MainActivity.onCreate}.
	\]

	Let $W$ be the set of \emph{WebView allocation sites} of~$a$, one for each \texttt{new WebView(\ldots)} statement, and let $\mathrm{pt}(r) \subseteq W$ be the allocation sites that a reference $r$ may denote. Each WebView $w \in W$ then owns the sites that may act on it:
	\[
	  C_w = \{\, c \in C \mid w \in \mathrm{pt}(\mathrm{wvRecv}(c)) \,\}.
	\]
\end{definition}

The alias information is obtained through light-weight one-object sensitive~\cite{object-sensitive-milanova}, flow-insensitive, field-sensitive points-to analysis.

Static analysis rarely recovers a string in full, so it records whatever part it does know as a \emph{pattern}~$\hat{s}$: literal text with wildcards~$\star$ standing for the parts it could not determine. The URL assembled on \cref{l2} of \Cref{lst:running}, for instance, comes out as \texttt{file:///android\_asset/}$\star$\texttt{.html}. A concrete string~$v$ is \emph{compatible} with~$\hat{s}$, written $\mathrm{compatible}(v,\hat{s})$, if substituting text for the wildcards of~$\hat{s}$ can produce~$v$. The pattern above is compatible with \texttt{file:///android\_asset/app.html}, the value this app in fact loads. A pattern with no wildcard is \emph{exact}: exactly one string is compatible with it. A bare wildcard, written~$\top$, is \emph{unknown}: every string is compatible with it. A pattern that mixes text and wildcards, like the one above, is \emph{partial}.

\begin{definition}[Resolution Map]\label{def:resmap}
	A \emph{resolution map} $\Phi$ gives, for each communication site, the candidate values recovered for its argument: $\Phi(c) \subseteq \mathit{String} \cup \{\top\}$. A site whose value is not yet settled is kept in the \emph{unresolved set} $U$, paired with the pattern Step~1 recovered for it.
\end{definition}

\begin{algorithm}[!b]
	\setlength{\belowcaptionskip}{-12pt}
	\small
	\DontPrintSemicolon
	\SetKwFunction{FSR}{StaticResolve}
	\SetKwFunction{FDyn}{DynamicRefine}
	\SetKwBlock{StepA}{Step 1: static value recovery}{}
	\SetKwBlock{StepB}{Step 2: guided dynamic refinement}{}
	\SetKwBlock{StepC}{Step 3: conservative over-approximation \textnormal{---} Def.~\ref{def:fallback}}{}
	\SetKwBlock{StepD}{Summary assembly}{}
	\KwIn{APK $a$}
	\KwOut{Per-WebView summary map $\Sigma = \{\sigma_w \mid w \in W\}$}
	$W \leftarrow$ WebView allocation sites in $a$;\quad $W_{\mathit{bnd}} \leftarrow$ bundled \texttt{.html}/\texttt{.js} assets \tcp*{Defs.~\ref{def:sites},~\ref{def:fallback}}
	$C \leftarrow$ communication sites in $a$;\quad $\Phi \leftarrow \emptyset$;\quad $U \leftarrow \emptyset$ \tcp*{nothing recovered yet}
	\StepA{
		\ForEach{$c \in C$}{
			$\hat{s} \leftarrow$ \FSR{$\mathrm{arg}(c)$} \tcp*{one pattern, by slicing back from $\mathrm{arg}(c)$}
			\lIf(\tcp*[f]{nothing left to determine}){$\hat{s}$ \textnormal{is exact, say} $v$}{$\Phi(c) \leftarrow \{v\}$}
			\lElse{$U \leftarrow U \cup \{(c,\hat{s})\}$ \tcp*[f]{keep the pattern with the site}}
		}
	}
	\StepB{
		$\Phi, U \leftarrow$ \FDyn{$a$, $\Phi$, $U$} \tcp*{\Cref{alg:dynrefine}}
	}
	\StepC{
		\ForEach{$(c,\hat{s}) \in U$}{
			\Switch{$\wkind(c)$}{
				\lCase{$\mathsf{bridge}$}{$\Phi(c) \leftarrow \mathrm{bridgeCandidates}(W_{\mathit{bnd}})$}
				\lCase{$\mathsf{load}$}{$\Phi(c) \leftarrow \{s \in W_{\mathit{bnd}} \mid \mathrm{compatible}(s,\hat{s})\}$}
				\lCase{$\mathsf{inject}$}{$\Phi(c) \leftarrow \{\top\}$ \tcp*[f]{any bridge call is possible}}
			}
		}
	}
	\StepD{
		\lForEach{$w \in W$}{$\Sigma[w] \leftarrow (\mathcal{I}_w, \mathcal{S}_w)$ \tcp*[f]{Def.~\ref{def:partition}}}
	}
	\Return{$\Sigma$}
	\caption{Phase 1: Per-WebView Summary Construction. Steps~1 and~3 over-approximate; Step~2 narrows.}
	\label{alg:summary}
\end{algorithm}

\begin{algorithm}[tbp]
	\small
	\DontPrintSemicolon
	\KwIn{APK $a$, resolution map $\Phi$, unresolved set $U$}
	\KwOut{Refined map $\Phi$, residual unresolved set $U$}
	instrument the WebView calls of \Cref{def:apis}\;
	execute the instrumented app with guided exploration\;
	$E \leftarrow$ collected runtime events $\langle \tau, \kappa, v \rangle$\;
	\ForEach{$(c,\hat{s}) \in U$}{
		$V \leftarrow \{\, v \mid \langle \tau,\kappa,v \rangle \in E,\; \tau = \mathrm{wvApi}(c),\; \mathrm{ctxCompatible}(\kappa,c) \,\}$\;
		\lIf{$\hat{s}$ \textnormal{is partial}}{$V \leftarrow \{\, v \in V \mid \mathrm{compatible}(v,\hat{s}) \,\}$}
		\If{$V \neq \emptyset$}{
			$\Phi(c) \leftarrow V$\;
			$U \leftarrow U \setminus \{(c,\hat{s})\}$\;
		}
	}
	\Return{$\Phi$, $U$}
	\caption{Step 2: Guided Dynamic Refinement}
	\label{alg:dynrefine}
\end{algorithm}

\noindent\textbf{Step~1: Static value recovery.}
The static resolver is selective. It starts from communication-relevant WebView API calls and recovers only the values that affect Java--JavaScript communication. For each communication site $c \in C$, it starts from $\mathrm{arg}(c)$ and traces the definitions that may reach that argument. Along this slice, it recovers string values through local assignments, recoverable field reads, common string constructors, and bounded flows through helper-method parameters and return values. When the slice reaches a bound interface object, it also collects updates on that object that may affect Java--JavaScript communication. These updates are kept as abstract field assignments and later become part of the WebView summary. It does not model full heap analysis, reflection, or arbitrary string solving. Step~1 is therefore focused on WebView communication, not on general-purpose string analysis.

For each site $c$, the analysis computes one pattern $\hat{s}$. If it is exact, say \texttt{"bridge"}, the value is settled and we set $\Phi(c) \leftarrow \{\texttt{"bridge"}\}$; such sites need no further processing. Otherwise $\hat{s}$ is partial, when part of the string is known but at least one component is not, or $\top$, when the value depends on runtime inputs, reflection, or flows the resolver does not model. We keep the unresolved part explicit rather than forcing an imprecise concrete value, and record the site as the pair $(c,\hat{s})$ in the unresolved set $U$, which is passed to Step~2.

\noindent\textbf{Step~2: Guided dynamic refinement.}
Step~1 leaves some values unresolved. These sites depend on values computed only at runtime. We therefore rewrite the app and insert logging calls around communication-relevant WebView APIs. During refinement, we consult only the runtime events that are relevant to unresolved sites. We instrument \texttt{loadUrl}, \texttt{loadData}, \texttt{loadDataWithBaseURL}, \texttt{evaluateJavascript}, and \texttt{addJavascriptInterface}. The logger records the API type, the observed string value, and the enclosing native-side context. At \texttt{addJavascriptInterface}, it also records the interface name and the runtime class of the bound interface object. At method entry, it emits the current context once and records nearby static URL or interface hints when they are available.

We then execute the instrumented app in a guided way. The static phase tells us which activities and fragments create the relevant WebView instances. We first try to launch these candidates directly. Next, we issue guided \texttt{VIEW} intents and perform a small deterministic sequence of UI actions. Finally, we use a seeded Monkey run as a fallback. The probe is not meant to cover the whole app. Its role is to expose unresolved WebView communication sites and record the concrete values that reach them.

Each runtime observation yields an event $\langle \tau, \kappa, v \rangle$, where $\tau$ is the API type, $\kappa$ is the native-side class observed at the logging point, and $v$ is the observed value. We use this information to match the event back to unresolved sites in the same WebView summary. In our implementation, $\mathrm{ctxCompatible}(\kappa,c)$ is checked first with the full native-side class name and then, when needed, with its simple name or source hint. If Step~1 has already recovered a partial value, we use it as an extra compatibility check. \Cref{alg:dynrefine} summarizes this refinement step.

For each $(c,\hat{s}) \in U$, we collect the observed values in a set $V$. If $\hat{s}$ is partial, we discard from $V$ the values that do not match the structure already recovered statically, keeping $\{\, v \in V \mid \mathrm{compatible}(v,\hat{s}) \,\}$. If $V \neq \emptyset$, we refine the map by setting $\Phi(c) \leftarrow V$ and remove $(c,\hat{s})$ from $U$. The remaining sites proceed to Step~3. In this way, the dynamic phase refines the WebView summary without changing its structure.

Steps~1 and~3 over-approximate. Step~2 does the opposite. Replacing $\top$ by a finite set of observed values is a narrowing, and it is justified only for the executions we exercised. We make this choice deliberately, since an unresolved site carries no usable information while an observed one does. \Cref{sec:soundness} states the resulting guarantee precisely.

\noindent\textbf{Step~3: Conservative over-approximation.}
After Steps~1 and~2, some sites may still remain unresolved, i.e., they could not be resolved statically or from observed executions. For them, we apply a conservative fallback that depends on the API type

\begin{definition}[Fallback Rules]\label{def:fallback}
	Let $W_{\mathit{bnd}}$ be the \texttt{.html} and \texttt{.js} files shipped inside the APK. For each $(c,\hat{s}) \in U$ remaining after Steps~1 and~2, we set
	\[
	\begingroup
	\setlength{\arraycolsep}{2pt}
	\Phi(c) \leftarrow
	\left\{
	\begin{array}{@{}ll@{}}
		\mathrm{bridgeCandidates}(W_{\mathit{bnd}})
		& \text{if } \wkind(c)=\mathsf{bridge},\\[0.35ex]
		\{s \in W_{\mathit{bnd}} \mid \mathrm{compatible}(s,\hat{s})\}
		& \text{if } \wkind(c)=\mathsf{load},\\[0.35ex]
		\{\top\}
		& \text{if } \wkind(c)=\mathsf{inject},
	\end{array}
	\right.
	\endgroup
	\]
	Here $\mathrm{bridgeCandidates}(W_{\mathit{bnd}})$ is the set of identifiers that occur as the receiver of a method call in a bundled asset.
\end{definition}

\noindent
The first rule scans the app's bundled assets for interface-object uses, since the JavaScript code that invokes bridge methods is often shipped with the app itself. An interface name that stays $\top$ is not usable, because the cross-language edges of \Cref{sec:callgraph} reach a native method only through a concrete name. The second rule assumes that the WebView may load any bundled artifact that matches the partial information recovered in Step~1; if nothing is known statically, it may load any bundled artifact. The third rule records that an unresolved injected script may invoke any registered bridge method, which \Cref{sec:callgraph} handles through a single unknown-JavaScript node.

\paragraph{Summary assembly.}
Once $\Phi$ has been populated, we construct the summary for each WebView allocation site $w$. It records the JavaScript artifacts associated with $w$ and one interface binding for each object that may be bound to the same WebView. Each binding fixes the interface name and the bound object, and thereby the bridge methods exposed through that object and the field updates on that object that may affect Java--JavaScript communication. It also records whether any artifact of $w$ is unavailable for analysis, which is the only distinction the construction needs to draw among unavailable artifacts, because a remote page and an unresolved script are equally opaque to a static JavaScript analyzer.

\begin{definition}[Per-WebView Summary]\label{def:partition}
	The summary of a WebView $w$ has two components, $\sigma_w = (\mathcal{I}_w, \mathcal{S}_w)$: the objects JavaScript can reach through $w$, and the JavaScript artifacts that may run in it.

	A bridge site registers an object under a name, so the bindings of $w$ are
	\begin{equation}\label{eq:bridges}
		\mathcal{I}_w = \bigl\{(o, n) \mid c \in C_w,\; \wkind(c) = \mathsf{bridge},\;
		  o \in \mathrm{pt}(\mathrm{bridgedObj}(c)),\; n \in \Phi(c) \bigr\}.
	\end{equation}
	What else a binding contributes follows from $o$ and $w$, so we look it up rather than store it: $M(o)$ is the set of methods of $o$, declared or inherited, that carry \texttt{@JavascriptInterface}, and $F_w(o) = \mathit{fieldSummary}(o, C_w)$ collects the updates on $o$ recovered along communication-relevant slices rooted at sites in $C_w$, each a pair $(f,\hat{v})$ of a field and the pattern assigned to it.

	The second component collects the JavaScript artifacts associated with $w$:
	\begin{equation}\label{eq:scripts}
		\mathcal{S}_w = \bigl\{\mathit{classify}(\mathrm{wvApi}(c), s) \mid c \in C_w,\;
		  \wkind(c) \in \{\mathsf{load},\mathsf{inject}\},\; s \in \Phi(c) \bigr\}.
	\end{equation}
	The function $\mathit{classify}(\tau, s)$ assigns each artifact to a simple content abstraction. It yields $\mathit{BundledScript}(s)$ for app-shipped or inline HTML and JavaScript. It yields $\mathit{RemoteScript}(s)$ for remote URLs. It yields $\mathit{InjectedScript}(s)$ for code injected through \texttt{evaluateJavascript} and for \texttt{loadUrl("javascript:...")}. If the artifact remains unresolved, it yields $\top$.
\end{definition}

\noindent

\subsection{Phase 2: Inter-Language Call Graph Construction}\label{sec:callgraph}

Phase~2 takes the per-WebView summaries from Phase~1 and constructs the inter-language call graph. By this point, the communication relation of each WebView has already been recovered. Each summary tells us which interface bindings are available in the WebView and which JavaScript artifacts may execute in the same communication context. Phase~2 then connects the native-side edges, the JavaScript-side edges, and the cross-language edges induced by the summary.

The main challenge is that these cross-language edges do not follow ordinary language-level dispatch. A JavaScript-side call such as \texttt{n.m(...)} reaches a native-side method only because the interface name $n$ was previously bound to an object whose exposed methods contain $m$. Standard call graph construction does not model this dispatch directly. We therefore synthesize a dedicated entry point for each WebView instance, inspired by FlowDroid's dummy-main abstraction~\cite{flowdroid}. \Cref{alg:callgraph} makes the relevant lifecycle and interaction behavior explicit for that WebView.

A \texttt{WebView} progresses through initialization, interface-object registration, content loading, interaction, and teardown. The first three stages are sequential. The analysis creates the WebView abstraction, enables JavaScript execution, introduces a representative object for each binding together with its summarized field state, and dispatches each JavaScript artifact in $\mathcal{S}_w$ according to its abstraction. The interaction stage is the critical part. Here, the native side and JavaScript side execute asynchronously and may interleave non-deterministically.

\begin{algorithm}[tbp]
	\small
	\DontPrintSemicolon
	\KwIn{Per-WebView summaries $\Sigma = \{\sigma_w = (\mathcal{I}_w, \mathcal{S}_w) \mid w \in W\}$}
	\KwOut{Synthetic entry points $\{\mathit{dummyMain}_w\}$}
	\ForEach{$w \in W$ \textnormal{with} $\mathcal{I}_w \neq \emptyset$}{
		\tcc{Stage 1: setup}
		create a fresh WebView abstraction $wv$ for $w$\;
		enable JavaScript execution on $wv$\;
		\tcc{Stage 2: registration}
		\ForEach{$(o,n) \in \mathcal{I}_w$}{
			introduce a representative native-side object for $o$\;
			apply the summarized updates $F_w(o)$ to $o$\;
			bind $n$ to $o$ on $wv$\;
		}
		\tcc{Stage 3: content loading}
		\lForEach{$s \in \mathcal{S}_w$}{dispatch $s$ into the execution context of $wv$}
		\tcc{Stage 4: interaction (non-deterministic)}
		\lWhile{$*$}{non-det.\ choose T1 $\mid$ T2 $\mid$ T3 $\mid$ T4 $\mid$ T5 \tcp*[f]{Def.~\ref{def:interaction}}}
		\tcc{Stage 5: teardown}
		terminate $wv$\;
	}
	\caption{Phase 2: Synthetic Entry Point Generation}
	\label{alg:callgraph}
\end{algorithm}

\begin{definition}[Non-deterministic Interaction Model]\label{def:interaction}
	After registration and content loading, the interaction phase is a non-deterministic loop over five transitions, where $(o,n)$ ranges over $\mathcal{I}_w$:
	\begin{center}
		\renewcommand{\arraystretch}{1.2}
		\begin{tabular}{@{}c p{5.6cm} l@{}}
			\toprule
			& \textbf{Transition} & \textbf{Direction} \\
			\midrule
			T1 & JavaScript side calls $n.m(*)$ for $m \in M(o)$ & JavaScript $\to$ Native \\
			T2 & native side calls $wv.\texttt{evaluateJavascript}(*,*)$ & Native $\to$ JavaScript \\
			T3 & native side calls $wv.\texttt{loadUrl}(*)$ or $wv.\texttt{loadData}(*)$ & Native $\to$ JavaScript \\
			T4 & apply a summarized update $o.f \leftarrow \hat{v}$ for $(f,\hat{v}) \in F_w(o)$ & Native writes \\
			T5 & JavaScript side calls an exposed getter $n.m()$, $m \in M(o)$, that reads $o.f$ for $(f,\_) \in F_w(o)$ & JavaScript reads (via getter) \\
			\bottomrule
		\end{tabular}
	\end{center}
\end{definition}

\noindent
T1 is unconditional. It applies even when $\mathcal{S}_w$ contains $\top$ or a $\mathit{RemoteScript}$. In such cases the JavaScript code is not known precisely, so we do not enumerate its control-flow graph. Instead we introduce an unknown-JavaScript abstraction, which may call any bridge method exposed through the same bindings. Transition T4 ranges over the field updates already summarized in Phase~1. We replay them flow-insensitively on the representative object so that the exposed getters called from JavaScript observe the same native-side state. JavaScript cannot access the fields of a bridge object directly, since WebView exposes only the methods annotated with \texttt{@JavascriptInterface}. T5 therefore reads a field only through an exposed getter; it is a special case of T1 that we list separately because it observes the updates replayed by T4. This preserves the native-side call edges needed for sound cross-language reasoning while keeping the construction modular.

The interaction loop is followed by a final teardown stage ($wv.\texttt{destroy}()$), which bounds the model's scope. Together, these stages give a per-instance abstraction of WebView that is sound when the communication relation is fully recovered statically, and otherwise conservative but coverage-dependent, since some edges may be learned from dynamic executions. The behavior is precise enough for downstream information-flow analyses and avoids the spurious merging that arises in app-wide models.

\paragraph{Cross-language edge semantics.}
The summary $\sigma_w$ and the interaction model together define a per-instance call graph $G_w = (N_H^w \cup N_G^w,\; E_H^w \cup E_G^w \cup E_{\times}^w)$. Here $N_H^w$ and $E_H^w$ are the reachable native-side nodes and edges, and $N_G^w$ and $E_G^w$ are the JavaScript-side nodes and edges induced by $\mathcal{S}_w$, together with a summary node for unresolved JavaScript behavior. The set $E_{\times}^w$ contains the cross-language edges. All three components are scoped to the same WebView instance, so JavaScript loaded in one WebView cannot spuriously invoke bridges registered on another.

The cross-language edges serve three purposes. First, they model JavaScript-to-native dispatch, from a call $n.m(\vec{a})$ to the native-side method $o.m$, but only when $(o,n) \in \mathcal{I}_w$ and $m \in M(o)$. Second, they model native-to-JavaScript transfer for \texttt{evaluateJavascript} and \texttt{loadUrl("javascript:...")}. Third, they model flow-insensitive field edges, from the summarized native-side updates in $F_w(o)$ to the exposed getters that read the same field and return its value to JavaScript, following the peculiar semantics of WebView~\cite{ifc-analysis-issre}.

We further apply two conservative rules. First, JavaScript code may redefine an exposed method by assigning a new JavaScript function to it, so a call through $n.m$ is connected to both the original Java implementation and any JavaScript-defined replacement~\cite{chin2013vunerabilities,neugschwandtner2013exploitation}. Second, deleting an individual bridge method does not remove the underlying Java entry point in the Android runtime; only deletion of the entire bridge object (\texttt{delete}~$n$) removes those edges~\cite{ifc-analysis-issre}.

\subsection{Stub Generation with Type and Effect System}\label{sec:typesystem}

In order to analyse the JavaScript part of the application by itself, we first have to determine the information flow effects of the bridge function. From the previous stage we have the call-graph $G$ and use it to compute the information flow labels, treating the bridge functions as entry-points. In principle, this could be achieved through a backward dataflow analysis; we instead use a type-and-effect system, which yields a compositional summary for each method that can be translated directly into a stub. In the following, we present the type-and-effect~\cite{ppa-nielson} system which computes each bridge method's confidentiality and integrity effects over the inter-language call graph $G$. The effects are translated into stubs which are then used by the modular JavaScript analysis. 

The typing rules themselves follow the classic discipline of security type systems~\cite{volpano96}: labels are propagated through explicit flows, a program-counter label captures implicit flows, and subtyping relates labeled types. Such type systems assume a closed world, in which the entire program is known statically and no code is generated at run time. A hybrid app violates this assumption: the names under which bridge objects are exposed, the pages that are loaded, and the scripts that are injected are frequently computed only at run time. Our contribution therefore does not lie in the individual rules, but in making this discipline applicable to an open program. The per-WebView summaries of Phase~1, refined by dynamic resolution and completed by the conservative fallback, close the program: they fix the bridge bindings and JavaScript artifacts, and hence the inter-language call graph over which the type system operates. Conversely, the type system complements the dynamic phase: it computes effects for all paths of the resulting call graph, including those that the dynamic exploration never executed, and exports them as stubs, so that the guest side can be analysed by existing single-language analysers.

We first define \textsc{WV-PolyDSL}, which encodes the constructs required for the discussion. \Cref{fig:host-syntax} shows the syntax of host statements in \textsc{WV-PolyDSL}. Variables are denoted by $v$ and $v'$, $g$ denotes a field name, $\tau$ a type, $i$ an object identifier, and $\mathit{script}$ a dynamically evaluated code fragment. The language includes object creation ($v = \mathbf{new}\;\tau()$), assignment ($v = v'$), field writes ($v.g = v'$), field reads ($v = v'.g$), and method calls ($v = f(v_1,\ldots,v_n)$), where $f$ denotes a method; a call binds the formal parameters of $f$ to the actual arguments $v_1,\ldots,v_n$, executes $\mathit{body}(f)$, and stores the value returned by $f$ in $v$. The construct $\mathbf{register}(v,\text{``name''})$ registers a host-side object $v$ under the guest-side name ``name'', making it accessible from guest code. The construct $v = \mathbf{eval}(\mathit{script})$ models dynamic evaluation of a JavaScript code fragment whose result is stored in $v$; in the host-side type-and-effect system, we conservatively treat this construct as $\mathbf{skip}$, because the dynamically executed guest code is analyzed through the modular JavaScript phase rather than through additional host-side transfer rules. The statement $\mathbf{async\_update}(i.g, v)$ performs a non-blocking update of field $g$ of the object identified by $i$~\cite{ifc-analysis-issre}. The statement $\mathbf{return}\;v$ returns the value of $v$ to the caller of the enclosing method; for a bridge method, this is how host data is handed back to the guest. We model it as a write of $v$ to a distinguished location $\mathit{ret}$ of the enclosing method. To keep the semantics simple, we assume that $\mathbf{return}$ occurs only in tail position, i.e., no statement of the method executes after it (it may still end both branches of a final conditional), so writing $\mathit{ret}$ coincides with leaving the method. Early returns are eliminated beforehand by the standard transformation that stores the returned value in a fresh variable, sets a flag, and guards the remaining statements with that flag; since the flag becomes a guard, the dependence of the rest of the method on an early return is tracked through $\TSpc$ and $\TSpct$. Sequential composition ($s_1 ; s_2$), conditional execution ($\mathbf{if}\;v\;\mathbf{then}\;s_1\;\mathbf{else}\;s_2$), and looping ($\mathbf{while}\;v\;\mathbf{do}\;s$) complete the language with the standard semantics.

\begin{figure}
  {\small
\begin{equation}
\begin{aligned}
  s \;::=\;&
    v = \mathbf{new}\;\tau()
    \mid v = v'
    \mid v.g = v'
    \mid v = v'.g \mid v = f(v_1,\ldots,v_n) \mid s_1; s_2 \\
    \mid& \mathbf{register}(v, \text{``name''})
    \mid v = \mathbf{eval}(\mathit{script})  \mid \mathbf{async\_update}(i.g, v) \\
    \mid & \mathbf{if}~v~\mathbf{then}~s_1~\mathbf{else}~s_2 \mid \mathbf{while}~v~\mathbf{do}~s \mid \mathbf{return}~v \\
\end{aligned}
\end{equation}
}
\caption{ \textsc{WV-PolyDSL}: Host Syntax}
\label{fig:host-syntax}
\end{figure}

\paragraph{Running example.}
In \textsc{WV-PolyDSL}, the host side of \Cref{lst:running} consists of the method \textit{onCreate}, which registers the bridge and loads the page, and of the two bridge methods of \texttt{API}:
\[
\begin{array}{@{}l@{\quad}l@{}}
\textit{onCreate}(): & \mathit{api} = \mathbf{new}\;\texttt{API}();\;\; \mathbf{register}(\mathit{api}, \text{``bridge''});\;\; r = \mathbf{eval}(\texttt{app.html}) \\
\textit{getPassword}(): & t = \mathit{this}.\mathit{password};\;\; \mathbf{return}\;t \\
\textit{log}(\mathit{msg}): & u = \mathit{println}(\mathit{msg})
\end{array}
\]
The interface name ``bridge'' and the loaded page \texttt{app.html} are the values recovered for \texttt{prefix} and \texttt{url} in Phase~1, and loading the page is modeled by $\mathbf{eval}$. The guest side, \texttt{pwd = bridge.getPassword(); bridge.log(pwd)}, is not part of \textsc{WV-PolyDSL}; it is analyzed later by the JavaScript analyser.

Figure~\ref{fig:typerules} defines the typing rules of \textsc{WV-PolyDSL} that statically capture whether a bridged method may result in an information leak. A \emph{labeled type} $\tau ::= B^\ell$ pairs a base type ($B ::= \text{Int} \mid \text{String} \mid \text{Object} \mid \text{Unit}$) with a security label $\ell$. A \emph{method type} $\forall\vec{\alpha}.\,\sigma \mid C$ with $\sigma ::= (\tau_1, \ldots, \tau_n) \xrightarrow{\TSpc_f,\,\TSpctf,\,\varepsilon} \tau_r$ is a \emph{parametric summary} of a method $f$, computed once and independently of its callers. Its parameter types are $\tau_j = B_j^{\alpha_j}$, where the label variable $\alpha_j$ stands for the confidentiality label of the $j$-th actual argument; likewise, inside the body of $f$ each formal parameter $p_j$ depends on itself, $p_j \in \Psi(p_j)$, standing for the bridge parameters on which the $j$-th actual argument depends. Labels in the return type and the effect, the alarms in the effect, and the provenance of $\mathit{ret}_f$ may mention $\alpha_j$ and $p_j$, and $C$ collects the constraints that the body imposes on fields and that mention them, e.g., $\alpha_j \sqsubseteq \ell_g$ and $p_j \in \Psi(o.g)$ when the $j$-th parameter is written to the field $o.g$. These constraints are not solved inside $f$ but instantiated with the actual arguments at each call site (\textsc{T-Call}). The method type also annotates the signature with program-counter bounds $\TSpc_f$ and $\TSpctf$: the body of the method is typed under them, so the method may only be called in contexts whose program-counter label is at most $\TSpc_f$ and whose guards are included in $\TSpctf$. Typing judgments have the form $\TSpc;\TSpct;\Gamma;\Psi \vdash s : \langle c, \TSA \rangle$, where $\TSpc \in \{L,H\}$ is the \emph{program-counter label}, an upper bound on the confidentiality of the guards that control whether $s$ executes, $\TSpct$ is the \emph{integrity program-counter context}, the set of guard variables that control whether $s$ executes, $\Gamma$ is the typing environment, $\Psi$ is the integrity environment, $s$ is a statement, and $\langle c, \TSA\rangle$ is the effect of $s$. The type system is \emph{flow-insensitive}: $\Gamma$ and $\Psi$ are fixed for an entire method body, and the rules only constrain them instead of updating them from statement to statement. Each local variable, field, and return location $\mathit{ret}$ therefore receives a single label (and provenance set) that over-approximates all assignments to it within the method. A method is therefore typed once as a whole and yields a single effect summary, which is all that stub generation requires. The effect consists of a confidentiality component $c \in \{L,H\}$ and an integrity alarm predicate $\TSA$ for external-to-internal flows. The \emph{confidentiality} component states whether the method may return a confidential value: $c = H$ iff some $\mathbf{return}$ in $s$ may return a value that depends on $H$ data, either explicitly or because the return is controlled by an $H$ guard. The \emph{integrity} component~$\TSA$ is a set of alarms $\mathsf{alarm}(x,p)$, each recording that a value that may depend on the bridge parameter $p$ flows into the internal location $x$. Let $\TSPar$ be the set of formal parameters of bridge methods, i.e., of methods exposed via \texttt{addJavascriptInterface} (\texttt{@JavascriptInterface}); these are the external inputs controlled by the guest. The integrity environment $\Psi : \textit{Loc} \rightarrow \mathcal{P}(\TSPar)$ records, for each access path in $\textit{Loc}$, the bridge parameters on which its value may depend; a location is external iff $\Psi(x) \neq \emptyset$, and each bridge parameter depends on itself (\textsc{T-Register}). Recording the parameters themselves, rather than only whether a location is external, lets every alarm name the parameter responsible for the flow, which is exactly what stub generation needs. The \emph{internal locations} $\TSInt \subseteq \textit{Loc}$, whose integrity is protected, are the host fields and the parameters of sink APIs.

Subtyping on labeled types requires $B_1 = B_2$ and $\ell_1 \sqsubseteq \ell_2$ (\textsc{Sub-Label}); on method types it is contravariant in parameters and in the program-counter bound, and covariant in return type and effect (\textsc{Sub-Method}). The typing environment~$\Gamma$ maps local variables, fields, and the return location $\mathit{ret}$ of each method to labeled types, and methods to method types; the integrity environment~$\Psi$ maps the same locations to sets of bridge parameters. Security labels form a lattice ordered by $\sqsubseteq$ with join operator $\sqcup$. Sets of bridge parameters are ordered by inclusion, with union as join. Effects are ordered component-wise, $\langle c,\TSA\rangle \sqsubseteq \langle c',\TSA'\rangle$ iff $c \sqsubseteq c'$ and $\TSA \subseteq \TSA'$, and accumulate via~$\sqcup$ across sequencing, where alarm sets are combined by union. For a destination $x$ and a set of bridge parameters $P$, we write $\mathsf{alarms}(x,P) \triangleq \{\mathsf{alarm}(x,p) \mid p \in P\}$ if $x \in \TSInt$, and $\emptyset$ otherwise. For a set of guards $G$, we write $\Psi(G) \triangleq \bigcup_{y \in G} \Psi(y)$ for the bridge parameters on which the guards may depend.
Rule \textsc{Sub-Method} defines sub-typing for method types with contravariant parameters and program-counter bounds, covariant return types, and monotonic effects: $\tau'_j <: \tau_j$, $\TSpc' \sqsubseteq \TSpc_f$, $\TSpct' \subseteq \TSpctf$, $\tau_r <: \tau'_r$, and $\langle c,\TSA\rangle \sqsubseteq \langle c',\TSA'\rangle$. A method whose body is safe under a high program-counter label can thus be used wherever a lower bound is expected. The subsumption rule \textsc{T-Sub} lets a statement be typed with a larger effect, under a smaller program-counter label, and under fewer guards: over-approximating whether a value is returned confidentially or which alarms are raised is always safe, a statement that only writes locations at least as confidential as $\TSpc$ also satisfies the weaker requirement imposed by any $\TSpc' \sqsubseteq \TSpc$, and a statement typed as if more guards controlled it raises at least as many alarms. Every rule that moves a value from a source $y$ into a location $x$ imposes the confidentiality constraint $\ell_y \sqcup \TSpc \sqsubseteq \ell_x$, which rules out both explicit flows and implicit flows through the guards recorded in $\TSpc$. The rule also imposes the provenance constraint $\Psi(y) \cup \Psi(\TSpct) \subseteq \Psi(x)$, so $x$ records every bridge parameter on which the value, or a guard controlling the assignment, may depend. \textsc{T-New}, \textsc{T-Assign}, and \textsc{T-Read} follow this pattern and have the trivial effect $\langle L,\emptyset\rangle$. Implicit flows are captured because every assignment inside a branch or loop body is typed under the raised $\TSpc$ and the extended guard set $\TSpct$, so its target must be at least as confidential as the guard and its provenance includes that of the guard. \textsc{T-Eval} conservatively treats dynamic guest-code evaluation as $\mathbf{skip}$ at the host level, producing no additional effect. The injected script is therefore not analysed by the type system: its content is recovered in Phase~1, statically or from runtime observations (\cref{sec:stringres}), extracted as a JavaScript artifact of the WebView ($\mathit{InjectedScript}(s)$ in \cref{def:partition}), and analysed by the JavaScript analyser together with the stubs (\cref{sec:modular-js}). \textsc{T-Register} makes every formal parameter $p$ of every method of a registered bridge object depend on itself, $p \in \Psi(p)$. Fields constitute the internal host state that integrity protects. Besides propagating provenance, \textsc{T-Write} and \textsc{T-Async} raise $\mathsf{alarms}(x, \Psi(y) \cup \Psi(\TSpct))$ for the written field $x$: one alarm $\mathsf{alarm}(x,p)$ for every bridge parameter $p$ on which the written value, or a guard controlling the write, may depend. Since provenance is propagated through local variables, fields, and calls, the alarm names the bridge parameter even when its value reaches $x$ only indirectly, e.g., through a local variable or a callee. \textsc{T-Return} constrains $\mathit{ret}$ like an assignment and raises the confidentiality effect to $\ell_v \sqcup \TSpc$, since the returned value leaves the method; for a bridge method, a return of $H$ data (or a return under an $H$ guard) is exactly a confidentiality leak to the guest. \textsc{T-Seq}, \textsc{T-If}, and \textsc{T-While} combine effects compositionally. Sequential composition joins confidentiality levels and combines alarm predicates with disjunction. Conditionals and loops type their branches (resp.\ loop body) under the raised program-counter label $\TSpc \sqcup \ell_c$ and add their guard $v$ to the guard set $\TSpct$; the guard label does not enter the effect directly, but a $\mathbf{return}$ under an $H$ guard yields $c = H$ through $\TSpc$. All other rules have confidentiality component $L$, since only \textsc{T-Return} hands a value back to the caller. Rule \textsc{T-Call} types a call $v = f(v_1,\ldots,v_n)$, where $B^\ell \sqcup \ell' \triangleq B^{\ell \sqcup \ell'}$ raises the label of a type. It instantiates the parametric summary of $f$ with the substitution $\theta$, which maps each label variable $\alpha_j$ to the label of the $j$-th argument, raised by $\TSpc$, and each formal parameter $p_j$ to the bridge parameters on which the argument or a controlling guard depends. On alarms, $\theta$ acts as $\mathsf{alarm}(x,p_j)\theta \triangleq \mathsf{alarms}(x, p_j\theta)$, and it leaves bridge parameters of other methods unchanged. The instantiated field constraints $C_f\theta$ must hold, so that, e.g., an $H$ argument may be written by $f$ only to an $H$ field. The instantiated type $\sigma_f\theta$ must be a subtype of the type determined by the call site: the arguments' types, raised by $\TSpc$, the program-counter label $\TSpc$ and guard set $\TSpct$ of the call site as the bounds, and the type of the target $v$. When the derivation tree is built, this premise is discharged by \textsc{Sub-Method}, and its premises by \textsc{Sub-Label}: contravariance requires $\TSpc \sqsubseteq \TSpc_f$ and $\TSpct \subseteq \TSpctf$, covariance lets the instantiated return value flow into $v$, and effect monotonicity makes the call inherit the instantiated alarms $\TSA_f\theta$ of the callee, which name the bridge parameters on which the actual arguments depend. The bound rules out implicit flows through calls: if $f$ is called under a high guard, then $\TSpc_f = H$, so the body of $f$ is typed under $H$ and cannot write to low locations. Likewise, if $f$ is called under a guard $y$, then $y \in \TSpctf$, so every write of $f$ to an internal field $x$ raises $\mathsf{alarm}(x,p)$ for every $p \in \Psi(y)$. The premise $\TSpc \sqsubseteq \Gamma(v).\ell$ accounts for calls under a raised program-counter label, and the provenance constraint passes the instantiated provenance of $\mathit{ret}_f$ into $v$, together with that of the guards. The confidentiality component of the call is $L$, because the returned value is stored in $v$ rather than returned by the caller; if the caller later returns $v$, \textsc{T-Return} accounts for it. Because the summary of $f$ refers to the arguments only through $\alpha_j$ and $p_j$, it is computed once, and each call site obtains its own instantiation. Finally, \textsc{T-Sink} types a call to a sink API $f$, whose parameters $f.p_j$ are internal locations: it raises $\mathsf{alarms}(f.p_j, \Psi(v_j) \cup \Psi(\TSpct))$ for every argument, i.e., one alarm for every bridge parameter that may reach the sink.

\paragraph{Running example: typing.}
For the running example, the policy $\Gamma_0$ labels the field $\mathit{password}$ as $H$ (developer annotation, see below) and classifies $\mathit{println}$ as a sink API, so its parameter $\mathit{println}.p_1$ is an internal location. Both methods are bridge methods, invoked from the guest with bounds $\TSpc = L$ and $\TSpct = \emptyset$. For \textit{getPassword}, the local $t$ and the return location receive the label $H$:
{\small
\begin{mathpar}
\inferrule*[right=T-Seq]
{
  \inferrule*[right=T-Read]
  { \Gamma(\mathit{this}.\mathit{password}) = \text{String}^{H} \\ \Gamma(t) = \text{String}^{H} \\\\ H \sqcup L \sqsubseteq H \\ \Psi(\mathit{this}.\mathit{password}) \cup \Psi(\emptyset) \subseteq \Psi(t) }
  { \TSJudge[L;\emptyset]{\Gamma}{\Psi}{t = \mathit{this}.\mathit{password}}{\TSEff{L}{\emptyset}} }
  \\
  \inferrule*[right=T-Return]
  { \Gamma(t) = \text{String}^{H} \\ \Gamma(\mathit{ret}) = \text{String}^{H} \\\\ H \sqcup L \sqsubseteq H \\ \Psi(t) \cup \Psi(\emptyset) \subseteq \Psi(\mathit{ret}) }
  { \TSJudge[L;\emptyset]{\Gamma}{\Psi}{\mathbf{return}\;t}{\TSEff{H}{\emptyset}} }
}
{ \TSJudge[L;\emptyset]{\Gamma}{\Psi}{t = \mathit{this}.\mathit{password};\; \mathbf{return}\;t}{\TSEff{H}{\emptyset}} }
\end{mathpar}
}
\noindent
The confidentiality component is $H$ because \textsc{T-Return} returns the $H$-labeled value $t$. For \textit{log}, the formal parameter $\mathit{msg}$ has the label variable $\alpha_1$ and depends on itself, $\Psi(\mathit{msg}) = \{\mathit{msg}\}$; it reaches the sink:
{\small
\begin{mathpar}
\inferrule*[right=T-Sink]
{ \mathit{println} \in \mathit{Sinks} }
{ \TSJudge[L;\emptyset]{\Gamma}{\Psi}{u = \mathit{println}(\mathit{msg})}{\TSEff{L}{\TSAlarm{\mathit{println}.p_1}{\{\mathit{msg}\}}}} }
\end{mathpar}
}
\noindent
Since $\mathit{println}.p_1 \in \TSInt$, $\mathsf{alarms}(\mathit{println}.p_1, \{\mathit{msg}\}) = \{\mathsf{alarm}(\mathit{println}.p_1, \mathit{msg})\}$: the guest-controlled parameter $\mathit{msg}$ reaches an internal location.

To illustrate the instantiation of summaries, suppose that \textit{log} delegated to a helper, $\textit{log}(\mathit{msg}): u = \mathit{write}(\mathit{msg})$ with $\textit{write}(x): u' = \mathit{println}(x)$. Typing \textit{write} by \textsc{T-Sink} as above yields the parametric summary $\forall\alpha.\,\sigma_w \mid \emptyset$ with $\sigma_w = \TSMArrow{\text{String}^{\alpha}}{L,\emptyset}{L}{\{\mathsf{alarm}(\mathit{println}.p_1, x)\}}{\text{Unit}^L}$, whose alarm names the helper's own parameter $x$. At the call site in \textit{log}, \textsc{T-Call} instantiates it:
{\small
\begin{mathpar}
\inferrule*[right=T-Call]
{ \Gamma(\mathit{write}) = \forall\alpha.\,\sigma_w \mid \emptyset \\ \theta = [\,\alpha \mapsto \alpha_1 \sqcup L,\; x \mapsto \{\mathit{msg}\}\,] \\\\
  \sigma_w\theta <: \TSMArrow{\text{String}^{\alpha_1}}{L,\emptyset}{L}{\TSA}{\Gamma(u)} \\
  L \sqsubseteq \Gamma(u).\ell \\ \Psi(\mathit{ret}_{\mathit{write}})\theta \cup \Psi(\emptyset) \subseteq \Psi(u) }
{ \TSJudge[L;\emptyset]{\Gamma}{\Psi}{u = \mathit{write}(\mathit{msg})}{\TSEff{L}{\TSA}} }
\end{mathpar}
}
\noindent
By effect monotonicity in \textsc{Sub-Method}, $\TSA \supseteq \{\mathsf{alarm}(\mathit{println}.p_1, x)\}\theta = \mathsf{alarms}(\mathit{println}.p_1, \{\mathit{msg}\}) = \{\mathsf{alarm}(\mathit{println}.p_1, \mathit{msg})\}$, so \textit{log} obtains the same alarm as before, now naming its own bridge parameter $\mathit{msg}$. Finally, \textsc{T-Register} for $\mathbf{register}(\mathit{api}, \text{``bridge''})$ in \textit{onCreate} requires $\mathit{msg} \in \Psi(\mathit{msg})$, i.e., that the parameter of the exposed method \textit{log} is a bridge parameter, which holds.

\begin{figure*}[!tbp]

\centering

{
  \small
  \mprset{vskip=1pt}
\begin{mathpar}
\inferrule[Sub-Label]
{
B_1 = B_2 \\
\ell_1 \sqsubseteq \ell_2
}
{
B_1^{\ell_1} <: B_2^{\ell_2}
}

\inferrule[Sub-Method]
{
\forall j.\; \tau'_j <: \tau_j \\
\tau_r <: \tau'_r \\\\
\TSpc' \sqsubseteq \TSpc_f \\
\TSpct' \subseteq \TSpctf \\
\TSEff{c}{\TSA} \sqsubseteq \TSEff{c'}{\TSA'}
}
{
\TSMArrow{\vec{\tau}}{\TSpc_f,\TSpctf}{c}{\TSA}{\tau_r}
<:
\TSMArrow{\vec{\tau'}}{\TSpc',\TSpct'}{c'}{\TSA'}{\tau'_r}
}

\inferrule[T-Sub]
{
\TSJudge{\Gamma}{\Psi}{s}{\TSEff{c}{\TSA}} \\
\TSpc' \sqsubseteq \TSpc \\
\TSpct' \subseteq \TSpct \\
\TSEff{c}{\TSA} \sqsubseteq \TSEff{c'}{\TSA'}
}
{
\TSJudge[\TSpc';\TSpct']{\Gamma}{\Psi}{s}{\TSEff{c'}{\TSA'}}
}

\inferrule[T-New]
{
\tau = B^{\ell} \\
\Gamma(v) = B^{\ell_v} \\\\
\ell \sqcup \TSpc \sqsubseteq \ell_v \\
\Psi(\TSpct) \TSLeqT \Psi(v)
}
{
\TSJudge{\Gamma}{\Psi}{v = \mathbf{new}\;\tau()}{\TSEff{L}{\emptyset}}
}

\inferrule[T-Eval]
{
\Gamma(v) = B^{\ell_v}
}
{
\TSJudge{\Gamma}{\Psi}{v = \mathbf{eval}(\mathit{script})}{\TSEff{L}{\emptyset}}
}

\inferrule[T-Register]
{
\forall m \in \text{methods}(v).\;
\forall p \in \mathit{params}(m).\;
p \in \Psi(p)
}
{
\TSJudge{\Gamma}{\Psi}{\mathbf{register}(v,n)}{\TSEff{L}{\emptyset}}
}

\inferrule[T-Assign]
{
\Gamma(v') = B^{\ell_s} \\
\Gamma(v) = B^{\ell_v} \\\\
\ell_s \sqcup \TSpc \sqsubseteq \ell_v \\
\Psi(v') \TSJoinT \Psi(\TSpct) \TSLeqT \Psi(v)
}
{
\TSJudge{\Gamma}{\Psi}{v = v'}{\TSEff{L}{\emptyset}}
}

\inferrule[T-Read]
{
\Gamma(v'.g) = B^{\ell_g} \\
\Gamma(v) = B^{\ell_v} \\\\
\ell_g \sqcup \TSpc \sqsubseteq \ell_v \\
\Psi(v'.g) \TSJoinT \Psi(\TSpct) \TSLeqT \Psi(v)
}
{
\TSJudge{\Gamma}{\Psi}{v = v'.g}{\TSEff{L}{\emptyset}}
}

\inferrule[T-Write]
{
\Gamma(v') = B^{\ell_s} \\
\Gamma(v.g) = B^{\ell_g} \\\\
\ell_s \sqcup \TSpc \sqsubseteq \ell_g \\
\Psi(v') \TSJoinT \Psi(\TSpct) \TSLeqT \Psi(v.g)
}
{
\TSJudge{\Gamma}{\Psi}{v.g = v'}{\TSEff{L}{\TSAlarm{v.g}{\Psi(v') \cup \Psi(\TSpct)}}}
}

\inferrule[T-Async]
{
\Gamma(v) = B^{\ell_v} \\
\Gamma(i.g) = B^{\ell_g} \\\\
\ell_v \sqcup \TSpc \sqsubseteq \ell_g \\
\Psi(v) \TSJoinT \Psi(\TSpct) \TSLeqT \Psi(i.g)
}
{
\TSJudge{\Gamma}{\Psi}{\mathbf{async\_update}(i.g,v)}{\TSEff{L}{\TSAlarm{i.g}{\Psi(v) \cup \Psi(\TSpct)}}}
}

\inferrule[T-Return]
{
\Gamma(v) = B^{\ell_v} \\
\Gamma(\mathit{ret}) = B^{\ell_r} \\\\
\ell_v \sqcup \TSpc \sqsubseteq \ell_r \\
\Psi(v) \TSJoinT \Psi(\TSpct) \TSLeqT \Psi(\mathit{ret})
}
{
\TSJudge{\Gamma}{\Psi}{\mathbf{return}\;v}{\TSEff{\ell_v \sqcup \TSpc}{\emptyset}}
}

\inferrule[T-Seq]
{
\TSJudge{\Gamma}{\Psi}{s_1}{\TSEff{c_1}{\TSA_1}} \\
\TSJudge{\Gamma}{\Psi}{s_2}{\TSEff{c_2}{\TSA_2}}
}
{
\TSJudge{\Gamma}{\Psi}{s_1;\,s_2}{\TSEff{c_1 \sqcup c_2}{\TSA_1 \cup \TSA_2}}
}

\inferrule[T-If]
{
\Gamma(v) = B^{\ell_c} \\\\
\TSJudge[\TSpc \sqcup \ell_c;\, \TSpct \cup \{v\}]{\Gamma}{\Psi}{s_1}{\TSEff{c_1}{\TSA_1}} \\\\
\TSJudge[\TSpc \sqcup \ell_c;\, \TSpct \cup \{v\}]{\Gamma}{\Psi}{s_2}{\TSEff{c_2}{\TSA_2}}
}
{
\TSJudge{\Gamma}{\Psi}{\mathbf{if}\;v\;\mathbf{then}\;s_1\;\mathbf{else}\;s_2}{\TSEff{c_1 \sqcup c_2}{\TSA_1 \cup \TSA_2}}
}

\inferrule[T-While]
{
\Gamma(v) = B^{\ell_c} \\
\TSJudge[\TSpc \sqcup \ell_c;\, \TSpct \cup \{v\}]{\Gamma}{\Psi}{s}{\TSEff{c_s}{\TSA_s}}
}
{
\TSJudge{\Gamma}{\Psi}{\mathbf{while}\;v\;\mathbf{do}\;s}{\TSEff{c_s}{\TSA_s}}
}

\inferrule[T-Call]
{
\Gamma(f) = \forall\vec{\alpha}.\;\sigma_f \mid C_f \\
\theta = [\,\alpha_j \mapsto \Gamma(v_j).\ell \sqcup \TSpc,\; p_j \mapsto \Psi(v_j) \cup \Psi(\TSpct)\,]_j \\
C_f\theta
\\\\
\sigma_f\theta <:
\TSMArrow{\Gamma(v_1) \sqcup \TSpc,\ldots,\Gamma(v_n) \sqcup \TSpc}{\TSpc,\TSpct}{c}{\TSA}{\Gamma(v)}
\\\\
\TSpc \sqsubseteq \Gamma(v).\ell \\
\Psi(\mathit{ret}_f)\theta \TSJoinT \Psi(\TSpct) \TSLeqT \Psi(v)
}
{
\TSJudge{\Gamma}{\Psi}{v = f(v_1,\ldots,v_n)}{\TSEff{L}{\TSA}}
}

\inferrule[T-Sink]
{
f \in \mathit{Sinks}
}
{
\TSJudge{\Gamma}{\Psi}{v = f(v_1,\ldots,v_n)}{\TSEff{L}{\textstyle\bigcup_j \TSAlarm{f.p_j}{\Psi(v_j) \cup \Psi(\TSpct)}}}
}
\end{mathpar}
}

\caption{Typing rules for the type-and-effect system over \textsc{WV-PolyDSL}.}
\label{fig:typerules}

\end{figure*}

\Cref{alg:effectcomp} presents the interprocedural effect computation. It takes as input the call graph $G$ and a starting host method $m$. The output is a mapping $\mathcal{E}$ from methods to their computed effects. The algorithm initializes the typing environment $\Gamma_0$ and constructs the reachable call subgraph $G_m$ rooted at $m$. It initializes all method summaries to the least effect $\varepsilon_\bot = \langle L, \emptyset\rangle$. Then, it processes strongly connected components (SCCs) of $G_m$ in bottom-up topological order so that callees are analysed before their callers whenever the call relation is acyclic. For each SCC, it iteratively re-types each function body $\mathit{body}(f)$ using $\textsc{TypeStmt}$ under the current summary map $\mathcal{E}$ until no summary changes. In this way, call sites consult the already-computed effects of their callees, while recursive SCCs are handled by fixpoint iteration from $\varepsilon_\bot$. The program-counter bounds $\TSpc_f$ and $\TSpctf$ of a method, however, depend on its callers, which the bottom-up order visits later. The algorithm therefore starts from the bounds $(L,\emptyset)$ and, after each bottom-up pass, recomputes them from the call sites (line~\ref{alg:effectcomp:bounds}); the outer loop (line~\ref{alg:effectcomp:outer}) repeats until neither the summaries nor the bounds change. Since labels, guard sets, and alarm sets range over finite lattices and all updates are monotone, the iteration terminates. Because the type system is flow-insensitive, $\textsc{TypeStmt}$ does not follow the control flow of $\mathit{body}(f)$. It collects the constraints that the rules in \cref{fig:typerules} impose on $\Gamma$ and $\Psi$ for every statement of $\mathit{body}(f)$, typed under the bounds $\TSpc_f$ and $\TSpctf$ of $f$, with parameter labels $\alpha_j$ and parameter provenance $\{p_j\}$. It solves the constraints on the locals and $\mathit{ret}$ of $f$ for the least labels and provenance sets, expressed in terms of $\alpha_j$ and $p_j$, and returns the parametric summary $\forall\vec{\alpha}.\,\sigma_f \mid C_f$, where $C_f$ contains the remaining constraints on fields. Because callers only instantiate this summary, a single bottom-up pass in the topological order of the SCCs suffices for the flows of arguments and return values; the outer loop is needed only for the program-counter bounds. The bound $\TSpc_f$ is the join of the program-counter labels at all call sites of $f$, and $\TSpctf$ is the union of their guard sets; a bridge method is additionally invoked from the guest as an entry point, which contributes $L$ and $\emptyset$, joined with the bounds of any host-side call sites. Here, $\Gamma_0$ supplies the security policy of the host side: the initial labels of sources, which are labeled $H$, and the sink APIs, whose parameters are internal locations, so that a call to a sink API is typed by \textsc{T-Sink} and raises an alarm for every bridge parameter that may reach it. The policy is a parameter of the analysis. One option is the Android source and sink classification of \textsf{SuSi}~\cite{susi}, which labels the values returned by source APIs; another is to let developers annotate sensitive variables and fields directly, as for the field \texttt{password} in \Cref{lst:running}. The type of each callee $g$ carries its current summary $\mathcal{E}(g)$ as its effect. The effect of the whole body becomes the summary of $f$. The analysis is conservative with respect to the constructed inter-language call graph: it may over-approximate flows, for instance because a field receives a single label for all writes to it, but it does not miss flows represented in the call graph.

\paragraph{Running example: summaries.}
Neither bridge method calls another host method, so \Cref{alg:effectcomp} needs a single pass, and the bounds of both methods remain $(L,\emptyset)$. The resulting parametric summaries are
\[
\begin{array}{@{}l@{\;:\;}l@{}}
\textit{getPassword} & () \xrightarrow{L,\,\emptyset,\,\langle H,\,\emptyset\rangle} \text{String}^{H} \\
\textit{log} & \forall \alpha_1.\; (\text{String}^{\alpha_1}) \xrightarrow{L,\,\emptyset,\,\langle L,\,\{\mathsf{alarm}(\mathit{println}.p_1,\,\mathit{msg})\}\rangle} \text{Unit}^{L}
\end{array}
\]
with empty field constraints $C$, since neither method writes a field.

\subsection{Modular JavaScript Analysis}\label{sec:modular-js}
\begin{algorithm}[h]
	\SetAlgoLined
	\LinesNumbered
	\DontPrintSemicolon
	
	\SetKwInOut{Input}{Input}
	\SetKwInOut{Output}{Output}
	
	\Input{Call graph $G=(N,E)$; invoking host method $m$}
	\Output{Effect summaries $\mathcal{E}: \text{method}\rightarrow\langle c,\TSA\rangle$ and bounds $(\TSpc_f,\TSpctf)$}
	
	\BlankLine
	
	$\Gamma_0 \leftarrow$ initial typing environment\;\label{alg:effectcomp:init}
	$G_m \leftarrow$ subgraph of $G$ reachable from $m$\;
	
	\ForEach{$f \in G_m$}{
		$\mathcal{E}(f) \leftarrow \varepsilon_\bot$;\quad $(\TSpc_f, \TSpctf) \leftarrow (L, \emptyset)$\;\label{alg:effectcomp:initeffects}
	}
	\Repeat{no summary and no bound changes}{\label{alg:effectcomp:outer}
		\ForEach{SCC $C$ of $G_m$ in bottom-up topological order}{\label{alg:effectcomp:scc}
			\Repeat{no summary changes}{\label{alg:effectcomp:fixpoint}
				\ForEach{$f \in C$}{
					$\varepsilon \leftarrow \textsc{TypeStmt}(\mathit{body}(f),\Gamma_0,\TSpc_f,\TSpctf,\mathcal{E})$\;\label{alg:effectcomp:typestmt}
					$\mathcal{E}(f) \leftarrow \varepsilon$\;\label{alg:effectcomp:update}
				}
			}
		}
		\ForEach{$f \in G_m$}{\label{alg:effectcomp:bounds}
			$(\TSpc_f, \TSpctf) \leftarrow$ join of $(\TSpc, \TSpct)$ over the call sites of $f$ in $G_m$\;
		}
	}
	\Return{$\mathcal{E}$}
	
	\caption{Interprocedural Effect Computation}
	\label{alg:effectcomp}
\end{algorithm}
\approach generates the stub for each bridge method. For each bridge method, it instantiates the summary for an invocation from the guest, mapping each $\alpha_j$ to $L$ and each $p_j$ to $\{p_j\}$, i.e., the arguments are public guest data and the bridge parameters themselves; for the resulting effect $\varepsilon=\langle c,\TSA\rangle$, it emits a stub object that mirrors the bridge API. For each formal parameter $p_j$, it emits a sink marker \texttt{\_\_leak\_value\_\_(}$p_j$\texttt{)} iff $\mathsf{alarm}(x,p_j) \in \TSA$ for some internal location $x$, i.e., the parameter $p_j$ may flow, explicitly or through a guard, into the internal location $x$ (a host field or a sink parameter). Correspondingly, it emits a source marker (\texttt{\_\_sensitive\_value\_\_()}) as the return value whenever $c=H$. The generated JavaScript precisely exposes the confidentiality and integrity effects without requiring the Java implementation. Guest-side redefinition and deletion of bridge methods are not modeled in \textsc{WV-PolyDSL}, because they change only guest code. As described in \cref{sec:callgraph}, a call $n.m$ is connected both to the original Java method and to any JavaScript replacement, and deleting an individual bridge method does not remove the underlying Java entry point. The stub therefore always carries the summary of the Java method, whereas a JavaScript replacement is ordinary guest code that the JavaScript analyser sees alongside the stub. The instrumented code (the original JavaScript code, including the scripts injected via \texttt{evaluateJavascript} and \texttt{loadUrl("javascript:...")} that Phase~1 recovered, prefixed with the instrumentation) is then provided to any off-the-shelf JavaScript analyser. On the guest side, each JavaScript analyser brings its own policy of sources and sinks; \approach{} adds \texttt{\_\_sensitive\_value\_\_()} as a source and \texttt{\_\_leak\_value\_\_()} as a sink to that policy. In this way, standard single-language tools detect every boundary-relevant flow, enabling a modular analysis. %

\paragraph{Running example: stubs.}
Instantiating the summaries for an invocation from the guest ($\alpha_1 \mapsto L$, $\mathit{msg} \mapsto \{\mathit{msg}\}$) leaves them unchanged. Since \textit{getPassword} has $c = H$, its stub returns a source marker; since the summary of \textit{log} contains $\mathsf{alarm}(\mathit{println}.p_1, \mathit{msg})$, its stub passes $\mathit{msg}$ to a sink marker:
\begin{lstlisting}[language=JavaScript, numbers=none]
var bridge = {
  getPassword: function () { return __sensitive_value__(); },
  log: function (msg) { __leak_value__(msg); }
};
var pwd = bridge.getPassword();   // original guest code (app.html)
bridge.log(pwd);
\end{lstlisting}
\noindent
A JavaScript analyser that treats \texttt{\_\_sensitive\_value\_\_()} as a source and \texttt{\_\_leak\_value\_\_()} as a sink now reports the flow from \texttt{getPassword} through \texttt{pwd} to \texttt{log}, i.e., the leak of the password to the public log, without analysing any Java code.

\providecommand{\LEquiv}[1]{\approx^{L,c}_{#1}}
\subsection{Soundness}\label{sec:soundness}

\subsubsection{Soundness of Interlanguage Callgraph}
Phase~2 inherits its call-graph edges from the resolution map $\Phi$ computed in Phase~1. We call an app $a$ \emph{statically resolved} when Step~1 already yields only exact values, i.e., $U = \emptyset$ after static value recovery. In that case, the inter-language call graph is built entirely from statically recovered interface names, bridge objects, field summaries, and JavaScript artifacts.

\begin{theorem}[Call-Graph Soundness for Statically Resolved Apps]\label{thm:cg-sound}
Let $a$ be statically resolved, let $w$ be a WebView allocation site of $a$, and let $G_w$ be the per-WebView inter-language call graph generated from $\Sigma[w]$. Then every concrete bridge registration, content-loading step, JavaScript-to-native dispatch, native-to-JavaScript transfer, and summarized field access reachable at runtime through the WebView instance represented by $w$ is contained in~$G_w$.
\end{theorem}

\noindent
Every cross-language edge in Phase~2 comes from three ingredients recovered in Phase~1: (i) concrete bridge registrations, which determine the interface-summary environment $\mathcal{I}_w$; (ii) concrete content-loading operations, which determine $\mathcal{S}_w$; and (iii) the summarized bridge fields attached to each registered object. If Step~1 resolves all communication-relevant values exactly, then these ingredients are fixed statically.

Once $\mathcal{I}_w$ and $\mathcal{S}_w$ are exact, the rest of the construction is syntactic. Equations~\ref{eq:bridges} and~\ref{eq:scripts} enumerate the exact interface-summary and script components of $\sigma_w$, and \Cref{def:interaction} adds every transition by which the two sides may interact: bridge registration and loading, JavaScript-to-native calls, native-side callbacks into JavaScript, and summarized read/write effects on exposed bridge state. Therefore every concrete runtime communication event has a corresponding abstract edge in $G_w$. The graph may still over-approximate JavaScript control flow or merge feasible interaction orders, but it does not miss any bridge-level communication through $w$.

If Step~1 leaves a communication site unresolved, the pipeline delegates to Step~2, which refines $\Phi$ with values observed during execution. This improves precision, but the guarantee changes: a dynamically recovered interface name, URL, or script fragment is justified only for the explored executions. A different execution may expose a previously unseen registration or content-loading behavior, and those edges need not appear in the graph.

Step~3 addresses this loss of coverage with a conservative fallback for sites that remain unresolved after dynamic analysis. The fallback does not guess one behavior. It replaces an unknown with a superset of behaviors that the graph construction must consider. This makes the over-approximation sound. The construction of $\sigma_w$ is monotone in $\Phi$: enlarging $\Phi(c)$ can only add interface bindings, artifacts, or edges, never remove one. So if the fallback maps an unresolved site to a set containing every behavior still compatible with the static evidence, the resulting graph contains every concrete edge induced by that site.

The three fallback cases in Definition~\ref{def:fallback} all follow this pattern. For an unresolved bridge registration, $\mathrm{bridgeCandidates}(W_{\mathit{bnd}})$ collects every bridge interface referenced by the app's bundled artifacts, so the analysis assumes any bundled bridge use may occur. For an unresolved content-loading site, the set $\{s \in W_{\mathit{bnd}} \mid \mathrm{compatible}(s,\hat{s})\}$ contains every bundled artifact consistent with the partial string recovered statically, so any concrete bundled page or script matching that evidence remains represented. For an unresolved injected script, the fallback value $\top$ is interpreted by the unknown-JavaScript abstraction in \Cref{def:interaction}; transition~T1 then allows calls to any bridge method exposed through $\mathcal{I}_w$. Although precise guest-side control flow is lost, every bridge dispatch that such an unseen script could perform remains in the abstract call graph.

At the level of concrete executions, the argument is direct. Consider a runtime event originating at a site still unresolved after Steps~1 and~2. If it is a bridge registration, the fallback includes all bridge candidates compatible with the bundled JavaScript. Equation~\ref{eq:bridges} turns each candidate into an interface summary, so every later JavaScript-to-native call through that unresolved registration is matched by some abstract T1 edge. If it is a load operation, the concrete page or script must be one of the artifacts consistent with the partial static evidence. Step~3 inserts all such compatible artifacts into $\Phi(c)$, and Equation~\ref{eq:scripts} propagates them into $\mathcal{S}_w$. The abstract graph may contain extra pages or scripts, but it cannot omit the concrete one as long as it is consistent with the recovered evidence.

The injected-script case makes the point most clearly. If \texttt{evaluateJavascript} receives a string that the analysis cannot resolve, then no precise JavaScript control-flow graph is available, so the analysis uses $\top$ instead of inventing a script. In the interaction model, $\top$ means that the guest side may perform any bridge call allowed by $\mathcal{I}_w$. This is an over-approximation because many such calls may never occur, but it is sound because every call the unknown script could make is included. The same reasoning applies to summarized field accesses: if a field may be read or written along unresolved guest behavior, the abstract model keeps that possibility. Step~3 preserves soundness by turning missing knowledge into nondeterminism. It never resolves unknowns optimistically; it expands them into admissible behaviors and includes edges for all of them.

This conservative treatment does not make the dynamically refined part fully sound, because Step~2 may resolve a site only to the values that happened to be observed. The guarantee is therefore mixed. Static recovery yields genuine soundness. Residual unknowns handled conservatively remain safe. Dynamically sharpened sites are only as complete as the executions used to discover them. For this reason, the overall call-graph generation is \emph{soundy}, in the sense of the ``soundiness'' position articulated by Livshits et al.~\cite{livshits2015soundiness}: it preserves a sound core where the static evidence suffices, and otherwise makes the remaining uncertainty explicit through conservative over-approximation rather than unsound optimistic resolution.

\subsubsection{Soundness of Type and Effect System}
The soundness of type and effect-system rests on the notion of non-interference for confidentiality and integrity properties. To motivate this, we first discuss the semantics of the \textsc{WV-PolyDSL} and then show the non-interference theorem for confidentiality and integrity properties.
\paragraph{Semantics.}
Security labels and effects are a static concern of the type system; the operational semantics is defined purely in terms of the concrete \emph{memory} $\mu:\mathit{Loc}\to\mathit{Val}$. Figure~\ref{fig:sem-rules} defines the big-step relation $s,\mu\Downarrow\mu'$, meaning that statement $s$ executed in memory $\mu$ terminates leaving memory $\mu'$. Branching (conditionals and loops) is resolved by testing the runtime value $\mu(v)$. We write $\mathit{Heap} \subseteq \mathit{Loc}$ for the field locations, $\mu|_X$ for the restriction of $\mu$ to $X$, and $\mu[\mu'|_X]$ for $\mu$ updated with the values of $\mu'$ on $X$. A call (\textsc{S-Call}) executes the callee in a fresh frame that shares only the heap with the caller and binds the formal parameters $p_j$ to the arguments; afterwards, the caller's local variables are restored, the callee's heap updates are kept, and the value of the callee's $\mathit{ret}$ is assigned to $v$. Thus parameters and locals of the callee never overwrite those of the caller, also under recursion. The semantics executes $\mathbf{async\_update}$ as an ordinary field write (\textsc{S-Async}) and does not model the interleaving of host and guest execution. This suffices for the type system: since it is flow-insensitive, each field receives a single label and provenance set that bound every value written to it, independently of the order in which the writes and the guest-side reads occur. The interleavings themselves are over-approximated by the call graph, whose interaction loop (\cref{def:interaction}) replays the summarized field updates flow-insensitively (T4) and lets the guest call the exposed getters that read these fields at any point (T5). Since such a getter is a bridge method, the value it returns is covered by its confidentiality effect: if the field may hold $H$ data, \textsc{T-Read} and \textsc{T-Return} give the getter $c = H$, and its stub emits the source marker.

The confidentiality label $\Gamma(x).\ell \in \{L,H\}$ of each location $x$ partitions memory into a \emph{low-confidentiality part} and a \emph{high-confidentiality part}:
\[
  \mu_L \;\triangleq\; \mu\big|_{\{\,x \mid \Gamma(x)=B^L\,\}}
  \qquad
  \mu_H \;\triangleq\; \mu\big|_{\{\,x \mid \Gamma(x)=B^H\,\}}.
\]
Two memories are \emph{low-confidentiality equivalent under $\Gamma$}, written $\mu_1\LEquiv{\Gamma}\mu_2$, iff they agree on every location whose confidentiality label is $L$:
\[
  \mu_1 \LEquiv{\Gamma} \mu_2
  \;\triangleq\;
  \forall x.\;\Gamma(x)=B^L \;\Rightarrow\; \mu_1(x)=\mu_2(x).
\]
Note that $\mu_H$ may differ between the two memories: high-confidentiality data is not observable by a low-confidentiality observer. Because the type system is flow-insensitive, $\Gamma$ (and likewise $\Psi$ below) is a single environment for the whole method, so these partitions do not change during execution. In the theorems, $\Gamma$ and $\Psi$ denote the union of the per-method environments for local variables and $\mathit{ret}$, the environment of the fields shared by all methods, and the method types; at each call, the callee's method type is instantiated by \textsc{T-Call}.

Dually, the integrity environment $\Psi$ records on which bridge parameters each location may depend. For a bridge parameter $p \in \TSPar$, two memories are \emph{$p$-integrity-equivalent under $\Psi$}, written $\mu_1 \approx^{p}_{\Psi} \mu_2$, iff they agree on every location that cannot depend on $p$:
\[
  \mu_1 \approx^{p}_{\Psi} \mu_2
  \;\triangleq\;
  \forall x.\;p \notin \Psi(x) \;\Rightarrow\; \mu_1(x)=\mu_2(x).
\]
Locations that may depend on $p$, in particular $p$ itself, may differ between the two memories, since $p$ models guest-controlled input. Thus, $p$-integrity-equivalence means that the two memories differ at most in what the guest controls through~$p$.

\begin{figure*}[!tbp]
\centering
{
  \mprset{vskip=1pt}
\begin{mathpar}

\inferrule[S-New]
{ o \text{ fresh} }
{ v = \mathbf{new}\;\tau(),\;\mu \;\Downarrow\; \mu[v\mapsto o] }

\inferrule[S-Eval]
{ }
{ v = \mathbf{eval}(\mathit{script}),\;\mu \;\Downarrow\; \mu }

\inferrule[S-Assign]
{ }
{ v = v',\;\mu \;\Downarrow\; \mu[v\mapsto\mu(v')] }

\inferrule[S-Read]
{ }
{ v = v'.g,\;\mu \;\Downarrow\; \mu[v\mapsto\mu(v'.g)] }

\inferrule[S-Write]
{ }
{ v.g = v',\;\mu \;\Downarrow\; \mu[v.g\mapsto\mu(v')] }

\inferrule[S-Return]
{ }
{ \mathbf{return}\;v,\;\mu \;\Downarrow\; \mu[\mathit{ret}\mapsto\mu(v)] }

\inferrule[S-Register]
{ }
{ \mathbf{register}(v,n),\;\mu \;\Downarrow\; \mu[n\mapsto\mu(v)] }

\inferrule[S-Async]
{ }
{ \mathbf{async\_update}(i.g,v),\;\mu \;\Downarrow\; \mu[i.g\mapsto\mu(v)] }

\inferrule[S-Seq]
{
s_1,\;\mu \;\Downarrow\; \mu_1 \\
s_2,\;\mu_1 \;\Downarrow\; \mu_2
}
{ s_1;\,s_2,\;\mu \;\Downarrow\; \mu_2 }

\inferrule[S-If-T]
{
\mu(v) \neq 0 \\
s_1,\;\mu \;\Downarrow\; \mu_1
}
{ \mathbf{if}\;v\;\mathbf{then}\;s_1\;\mathbf{else}\;s_2,\;\mu \;\Downarrow\; \mu_1 }

\inferrule[S-If-F]
{
\mu(v) = 0 \\
s_2,\;\mu \;\Downarrow\; \mu_2
}
{ \mathbf{if}\;v\;\mathbf{then}\;s_1\;\mathbf{else}\;s_2,\;\mu \;\Downarrow\; \mu_2 }

\inferrule[S-While-F]
{ \mu(v) = 0 }
{ \mathbf{while}\;v\;\mathbf{do}\;s,\;\mu \;\Downarrow\; \mu }

\inferrule[S-While-T]
{
\mu(v) \neq 0 \\
s,\;\mu \;\Downarrow\; \mu_1 \\
\mathbf{while}\;v\;\mathbf{do}\;s,\;\mu_1 \;\Downarrow\; \mu_2
}
{ \mathbf{while}\;v\;\mathbf{do}\;s,\;\mu \;\Downarrow\; \mu_2 }

\inferrule[S-Call]
{
\mu_f = \mu|_{\mathit{Heap}}[\,p_j \mapsto \mu(v_j)\,]_j \\
\mathit{body}(f),\;\mu_f \;\Downarrow\; \mu'
}
{ v = f(v_1,\ldots,v_n),\;\mu \;\Downarrow\; \mu[\mu'|_{\mathit{Heap}}][v\mapsto\mu'(\mathit{ret})] }

\end{mathpar}
}
\caption{Big-step operational semantics of \textsc{WV-PolyDSL} over memory $\mu$. The relation $s,\mu\Downarrow\mu'$ is purely store-transforming; security labels and effects are left to the type system. Conditionals (\textsc{S-If-T}/\textsc{S-If-F}) and loops (\textsc{S-While-F}/\textsc{S-While-T}) branch on the concrete runtime value $\mu(v)$.}
\label{fig:sem-rules}
\end{figure*}

The two equivalence relations above formalize what an observer is allowed to notice. For confidentiality, the observer can inspect only $L$-labeled locations. A confidentiality violation therefore occurs when two runs start from memories that agree on all $L$-labeled locations but differ on some $H$-labeled locations, and this hidden difference later causes some $L$-labeled output location to differ. For integrity, the protected part is the internal state. An integrity violation occurs when two runs differ only in the guest-controlled value of a bridge parameter $p$, and this difference later changes some internal location. The non-interference theorems below state that well-typed programs rule out exactly these two situations, except that integrity violations are allowed only when the effect system records the corresponding alarm.

Both theorems below are \emph{termination-insensitive}: they relate only executions that terminate ($s,\mu_i \Downarrow \mu_i'$), so information conveyed solely by (non-)termination is not covered.

\begin{theorem}[Confidentiality Non-Interference]\label{thm:ni}
Let $\TSpc;\TSpct;\Gamma;\Psi\vdash s : \langle c,\TSA\rangle$.
If\/ $\mu_1 \LEquiv{\Gamma} \mu_2$ and
$s,\mu_i \Downarrow \mu_i'$ for $i=1,2$,
then (i) $\mu_1' \LEquiv{\Gamma} \mu_2'$; and
(ii) if\/ $c = L$, then $\mu_1'(\mathit{ret}) = \mu_2'(\mathit{ret})$, i.e., the returned value does not depend on high-confidentiality data.
\end{theorem}

\noindent
Intuitively, the typing constraints ensure that no $H$-labeled data can influence any $L$-labeled location, so two executions that start from low-confidentiality-equivalent memories---which may have completely different high-confidentiality content---must produce low-confidentiality-equivalent output memories. The proof proceeds by structural induction on the derivation of $s,\mu\Downarrow\mu'$. Since \textsc{T-Sub} is the only rule that is not syntax-directed and it only lowers $\TSpc$ and enlarges the effect, we may assume without loss of generality that the typing derivation of $s$ ends in the syntax-directed rule for $s$, followed by at most one application of \textsc{T-Sub}; lowering $\TSpc$ only weakens the hypotheses we use, and enlarging the effect preserves the conclusions, because $c \sqsubseteq c'$ in part~(ii) and alarms are only added. The base cases (\textsc{S-New}, \textsc{S-Eval}, \textsc{S-Register}) either allocate a fresh object or leave $\mu_L$ unchanged. The assignment cases (\textsc{S-Assign}, \textsc{S-Read}, \textsc{S-Write}, \textsc{S-Async}, \textsc{S-Return}) write to location $x$ (for \textsc{S-Return}, $x=\mathit{ret}$); the constraint $\ell_y \sqcup \TSpc \sqsubseteq \ell_x$ of the corresponding typing rule requires every source label to be $L$ whenever the target label is $L$, so $\mu_1(x)=\mu_2(x)$ follows from the low-confidentiality-equivalence hypothesis. For the conditional cases, if the guard $v$ has label $L$, then $\mu_1(v)=\mu_2(v)$, both executions choose the same branch (\textsc{S-If-T} or \textsc{S-If-F}), and the induction hypothesis applies. If $v$ has label $H$, both branches are typed under $\TSpc = H$, so by the constraint $\ell_y \sqcup \TSpc \sqsubseteq \ell_x$ they only write $H$-labeled locations and leave $\mu_L$ unchanged, whichever branch is taken. The loop case (\textsc{S-While-T}/\textsc{S-While-F}) is analogous (for terminating executions), with the induction ordered by the height of the loop's big-step derivation, equivalently the number of remaining \textsc{S-While-T} unfoldings before reaching \textsc{S-While-F}; this is a well-founded measure on natural numbers. The call case (\textsc{S-Call}) follows from the induction hypothesis for $\mathit{body}(f)$: instantiating the summary of $f$ with $\theta$ replaces each $\alpha_j$ by the actual label of the $j$-th argument, so the instantiated constraints $C_f\theta$ and the subtyping premise of \textsc{T-Call} impose on the actual arguments and on $v$ the same label discipline that the body imposes on its parameters and $\mathit{ret}_f$, and since \textsc{Sub-Method} requires $\TSpc \sqsubseteq \TSpc_f$, the body of $f$ is typed under a program-counter label at least as high as that of the call site; in particular, a call under an $H$ guard only writes $H$-labeled locations. For part~(ii), $c = L$ means that every $\mathbf{return}\;v$ in $s$ is typed under $\TSpc = L$ with $\ell_v = L$; hence both executions reach the same returns, and the returned values agree by part~(i).

\begin{theorem}[Integrity Non-Interference]\label{thm:integrity-ni}
Let $\TSpc;\TSpct;\Gamma;\Psi\vdash s : \langle c,\TSA\rangle$ and $p \in \TSPar$.
If\/ $\mu_1 \approx^{p}_{\Psi} \mu_2$ and
$s,\mu_i \Downarrow \mu_i'$ for $i=1,2$, then:
(i) $\mu_1' \approx^{p}_{\Psi} \mu_2'$; and
(ii) for every internal location $x \in \TSInt$, if\/ $\mu_1(x)=\mu_2(x)$ but $\mu_1'(x) \neq \mu_2'(x)$, then $\mathsf{alarm}(x,p)\in\TSA$.
Hence, a bridge parameter $p$ can influence an internal location $x$, explicitly or through a guard, only if the effect records $\mathsf{alarm}(x,p)$.
\end{theorem}

\noindent
The proof is again by structural induction on the derivation of $s,\mu\Downarrow\mu'$, for a fixed $p \in \TSPar$. As before, \textsc{T-Sub} needs no separate case: dropping guards and adding alarms only weakens what has to be shown. For part~(i), every rule that writes a location $x$ from a source $y$ imposes $\Psi(y) \cup \Psi(\TSpct) \subseteq \Psi(x)$ (for \textsc{S-Call}, on the parameters and on $v$). If $p \notin \Psi(x)$, then $p$ is neither in $\Psi(y)$ nor in $\Psi(g)$ for any guard $g \in \TSpct$, so by the induction hypothesis the source and the guards agree in both executions, and so do the written values. The base cases (\textsc{S-New}, \textsc{S-Eval}, \textsc{S-Register}) are immediate. For conditionals, if $p \notin \Psi(v)$, both executions take the same branch and the induction hypothesis applies; otherwise both branches are typed with $v \in \TSpct$, so every location they write has $p$ in its provenance and is not constrained by $\approx^{p}_{\Psi}$. Loops are analogous, with the induction ordered as in the proof of \Cref{thm:ni}. For calls, the body of $f$ is typed under $\TSpctf \supseteq \TSpct$, the substitution $\theta$ maps each formal parameter $p_j$ to the provenance of the $j$-th argument, so the instantiated alarms $\TSA_f\theta$ and provenance of $\mathit{ret}_f$ refer to the caller's bridge parameters, and the frame discipline of \textsc{S-Call} restores the caller's locals. For part~(ii), let $x \in \TSInt$ with $\mu_1(x)=\mu_2(x)$ and $\mu_1'(x) \neq \mu_2'(x)$. Then some write to $x$ (\textsc{T-Write}, \textsc{T-Async}, or \textsc{T-Sink}) occurs in at least one execution whose source $y$ or a controlling guard differs between the two executions; by part~(i), applied to the execution up to that write, $p \in \Psi(y) \cup \Psi(\TSpct)$, so the rule emits $\mathsf{alarms}(x, \Psi(y) \cup \Psi(\TSpct)) \ni \mathsf{alarm}(x,p)$. \textsc{T-Call} inherits the alarms of the callee, and \textsc{T-Seq}, \textsc{T-If}, and \textsc{T-While} take the union of the alarms of their parts, so the alarm reaches~$\TSA$.

\subsubsection{Soundness of stub generation.}
Let $P$ be the original hybrid program and $\widehat{P}$ the instrumented JavaScript program obtained by replacing each bridge method $f$ with the stub generated from its summary $\mathcal{E}(f)=\langle c_f,\TSA_f\rangle$, instantiated for an invocation from the guest. The host-side theorems make the annotations of the stubs sound.

\begin{lemma}[Soundness of the stubs]\label{lem:stub}
Let $f$ be a bridge method with formal parameters $p_1,\ldots,p_n$.
(i) If the value that $f$ returns to the guest may depend on $H$-labeled host data, then $c_f = H$, and the stub of $f$ returns \texttt{\_\_sensitive\_value\_\_()}.
(ii) If the argument $p_j$ may influence an internal location $x \in \TSInt$, then $\mathsf{alarm}(x,p_j) \in \TSA_f$, and the stub of $f$ passes $p_j$ to \texttt{\_\_leak\_value\_\_}.
\end{lemma}

\noindent
Part~(i) follows from \Cref{thm:ni}~(ii) and part~(ii) from \Cref{thm:integrity-ni}~(ii); the stub then emits the markers by construction. The stubs may contain additional markers, since the type system over-approximates.

The guest side is analyzed by an off-the-shelf JavaScript analyzer, whose soundness we do not re-establish. Instead, the soundness of \approach{} is compositional: it follows from the soundness of the call graph, of the host-side annotations, and of the guest analyzer.

\begin{theorem}[Compositional Soundness]\label{thm:compositional}
Let $w$ be a WebView. If (A1) the call graph $G_w$ is complete, (A2) the host policy $\Gamma_0$ is correct, and (A3) the guest analyzer is sound for its policy extended with the two markers, then \approach{} reports every cross-language flow through $w$, i.e., every flow from a confidential host value to a guest sink or an internal host location, and every flow from a guest source to an internal host location.
\end{theorem}

\noindent
Here, completeness means that $G_w$ contains every bridge binding, script, and bridge call that occurs at run time through $w$, and correctness of $\Gamma_0$ means that every confidential host value originates from an $H$-labeled location and every host sink is in $\mathit{Sinks}$. By (A1), every bridge method invoked at run time through $w$ appears in $G_w$ and thus has a stub in $\widehat{P}_w$, and every script that executes in $w$ is part of $\widehat{P}_w$ or is covered by the unknown-JavaScript abstraction of transition~T1. A flow from a confidential host value leaves the host through a return of a bridge method whose value depends on $H$ data, which by (A2) originates from an $H$-labeled location; by \Cref{lem:stub}~(i), this return is a source marker in $\widehat{P}_w$. A flow into an internal host location enters the host through a bridge argument $p_j$; by \Cref{lem:stub}~(ii), $p_j$ is passed to a sink marker. The part of the flow between the bridge boundary and the guest source or sink is guest code in $\widehat{P}_w$, so by (A3) the guest analyzer reports it. Assumption~(A1) holds for statically resolved apps by \Cref{thm:cg-sound}; for sites resolved dynamically in Step~2 it holds only for the explored executions, so the overall guarantee inherits the soundy character of the call graph. Host data spliced into injected scripts is excluded, as discussed in \cref{sec:limitations}.

\section{Evaluation}
\label{sec:evaluation}
We evaluate \approach{} on a large corpus of apps from F-Droid~\cite{fdroid} and investigate four research questions:

\begin{enumerate}[topsep=0em]
	\item[\textbf{RQ1}] How prevalent and diverse are bridge-enabled WebViews in practice?
	\item[\textbf{RQ2}] How much does the dynamic phase improve over static analysis?
	\item[\textbf{RQ3}] How does \approach{} compare against the closest baseline?
	\item[\textbf{RQ4}] How effective is the end-to-end pipeline at
detecting information flows?
\end{enumerate}

\subsection{Dataset and Experimental Setup}
\label{sec:dataset}

F-Droid provides a large corpus of open-source Android apps. Our selection of apps was based on two criteria. First, download the latest versions of the apps that use at least one instance of WebView, i.e., hybrid apps. Second, the apps should establish at least one active Java/Kotlin-to-JavaScript communication bridge, i.e., they should use at least one \texttt{addJavascriptInterface} call. After the first filter, we ended up with 4,126 apps; after the second, 338. For the resulting 338 apps, sizes are right-skewed, with a mean of 32.17 MB and a median of 13.65 MB (IQR: 5.10–39.79 MB), ranging from 0.06 MB to 436.60 MB. Table~\ref{tab:corpus-summary} summarizes the resulting corpus.

The static analysis phase extracts 1,047 bridge-binding entries from \texttt{addJavascriptInterface} call sites across these 338 apps. These bridge calls span 313 distinct initiating classes (classes that initiate WebView) and 288 distinct bridge classes (classes whose bridge object is shared with JavaScript), with an average of 2.99 exposed methods per bridge and a maximum of 62. At the string-resolution level, \approach produces 1,106 entries across 303 packages. Of the 444 entries that originate from the static phase alone, 162 are resolved precisely (confidence\,=\,1.0), 106 are partially characterized (confidence\,=\,0.3, where the call site and class are identified but argument values remain symbolic), and 176 are unresolved (confidence\,=\,0.0). This mix of quality levels underscores the need for a subsequent dynamic analysis phase.

For the dynamic phase, we instrument and execute all 338 APKs on a single Android emulator. Rather than attempting exhaustive GUI exploration, the dynamic phase follows a bounded, heuristic strategy designed to reach WebView-hosting screens quickly. For each app, we (i)~explicitly launch every WebView-relevant activity inferred from the manifest and the static analysis, (ii)~issue a bounded number of deep-link \texttt{VIEW} intents, (iii)~execute a deterministic sequence of key events, taps, and swipes, and (iv)~run a seeded Monkey phase (seed\,=\,42) as fallback exploration.
Each app is given a 60-second log-capture window, with 3-second inter-launch delays, 5~seconds of initialization time, and 5~seconds of post-interaction settling time. The Monkey phase generates 180 or 280 events at 150\,ms throttle, depending on whether WebView fragments were statically detected. Of the 338 apps, 309 completed successfully, yielding the runtime data analyzed in RQ2. The remaining 29 fail before producing any dynamic evidence, primarily due to repackaging errors, ABI mismatches, or SDK version constraints.

\paragraph{Performance.}
The static analysis phase took approximately 13.1 hours using five parallel workers, corresponding to a throughput of 25.8\,apps/hour. In the dynamic phase, each APK undergoes instrumentation, repackaging, installation on the emulator, automated exploration, and log collection. This resulted in approximately 37.5 hours, with a mean of 6.7 minutes per app. The end-to-end pipeline (static + dynamic) thus processes the entire corpus in just over two days on commodity hardware (a MacBook Pro 14-inch with 36 GB RAM and an Apple M3 Max processor).

\begin{table}[tbp]
\centering
\caption{Summary of the large-scale evaluation corpus.}
\label{tab:corpus-summary}
\begin{tabular}{lr}
\hline
\textbf{Metric} & \textbf{Value} \\
\hline
F-Droid apps after initial filtering & 4,126 \\
Apps with at least one \texttt{addJavascriptInterface} call & 338 \\
Bridge-binding records & 1,047 \\
Distinct initiating classes & 313 \\
Distinct bridge classes & 288 \\
Average exposed methods per bridge & 2.99 \\
Packages with string-resolution entries & 303 \\
String-resolution entries & 1,106 \\
Dynamic-analysis apps (successful / total) & 309 / 338 \\
\hline
Static phase wall-clock time (5 workers) & 13.1\,h \\
Dynamic phase wall-clock time (1 emulator) & 37.5\,h \\
Mean dynamic time per app & 6.7\,min \\
\hline
\end{tabular}
\end{table}

\subsection{RQ1: What does the bridge surface look like in real apps?}
\label{sec:rq1}

The Java-JavaScript bridge surface is substantial but unevenly distributed. At the per-app level, the median app registers two Java-JavaScript bridges, 251 apps register between two and five, 32 register between six and ten, and six apps register more than ten, with a maximum of 22. The bridge interface names are diverse: 170 distinct names appear across the corpus, the most common being \texttt{\_cordovaNative} (146~occurrences), \texttt{Android} (100), and \texttt{flutter\_inappwebview} (41).

Structurally, the bridge surface reflects substantial framework reuse across apps. Of the 313 initiating classes, 42 appear in more than one app, indicating repeated reuse of shared library code, and six appear in ten or more apps. The most recurrent are \nolinkurl{org.apache.cordova.engine.SystemWebViewEngine} (69~apps), \nolinkurl{com.getcapacitor.MessageHandler} (33), \nolinkurl{com.getcapacitor.plugin.CapacitorHttp} and \nolinkurl{com.getcapacitor.plugin.CapacitorCookies} (29 each), \nolinkurl{com.pichillilorenzo.flutter_inappwebview_android.webview.in_app_webview.InAppWebView} (20), and \nolinkurl{com.reactnativecommunity.webview.RNCWebView} (15). A similar reuse pattern appears among the exposed bridge classes: \nolinkurl{org.apache.cordova.engine.SystemExposedJsApi} alone occurs in 68~apps.

This observation has a direct methodological implication: counting class occurrences would overestimate diversity by repeatedly capturing the same library behavior. To avoid this, we report all metrics at the application and entry levels. Even so, the corpus still shows meaningful variation beyond the recurring Cordova, Capacitor, Flutter, and React Native infrastructures, with a long tail of app-specific bridge calls.

\begin{findingbox}
\textbf{Answer to RQ1.}
Bridge-enabled WebViews are common in real-world apps, and the overall bridge surface is broad. At the same time, much of this surface is shaped by reused framework code rather than independent implementations. As a result, application-level counts provide a more accurate picture than raw class-level counts.
\end{findingbox}
\subsection{RQ2: How much does guided dynamic refinement help?}
\label{sec:rq2}

\begin{table}[tbp]
	\centering
	\caption{Sample entries for \textsc{com.spisoft.quicknote}, illustrating the three resolution outcomes. S\,=\,static, D\,=\,dynamic.}
	\label{tab:sample-entries}
	\small
	\begin{tabular}{p{0.73\columnwidth}cc}
		\hline
		\textbf{Resolved target (abbreviated)} & \textbf{Phase} & \textbf{Conf.} \\
		\hline
		\texttt{javascript:compatibility\allowbreak{}.sendNextLargeDownload()} & S & 1.0 \\
		\texttt{UNRESOLVED\_LOW\_QUALITY\_LITERAL:\allowbreak{} loadNote} & S & 0.0 \\
		\texttt{javascript:window.writer\allowbreak{}.handleAction(\textquotesingle prefill\textquotesingle, \textquotesingle\textquotesingle)} & D & 1.0 \\
		\hline
	\end{tabular}
\end{table}

Across the 309 successful runs, the dynamic analysis phase observes 668 runtime WebView calls and retains 662 as usable. These are distributed across four API methods: \texttt{loadUrl} (394, 59.0\%), \texttt{evaluateJavascript} (246, 36.8\%), \texttt{loadDataWithBaseURL} (18, 2.7\%), and \texttt{loadData} (10, 1.5\%).

We noticed an interesting pattern when comparing the static and dynamic phases by method type. The 444 static entries come entirely from three methods: \texttt{loadUrl} (333), \texttt{loadDataWithBaseURL} (71), and \texttt{loadData} (40). In the current implementation, the static phase does not produce any \texttt{evaluateJavascript} entries. The dynamic phase contributes all 246 \texttt{evaluateJavascript}-backed entries. In practice, many \texttt{evaluateJavascript} arguments are constructed at runtime through callback dispatch or framework-level plumbing, which makes them difficult to resolve statically. The dynamic phase, therefore, complements the static analysis by capturing these targets.

Dynamic analysis substantially improves runtime WebView resolution. The static analysis phase produces 444 string-resolution entries (URLs/JavaScript Code). The dynamic phase then adds 577 new entries and upgrades 85 existing ones, bringing the final total to 1,106. All 662 dynamic entries are resolved by construction, since the target values are observed at runtime. Combined with the 162 statically resolved entries, 824 of the 1,106 entries (74.5\%) carry a concrete resolution. Table~\ref{tab:dynamic-summary} summarizes the key metrics.

Of the 309 successful runs, 211 apps (68.3\%) produce at least one usable runtime entry, and 30 apps resolve at least one previously unknown target. Notably, 155 apps have no string-resolution entries (URL/JavaScript code) from the static phase alone; i.e., the analysis detects \texttt{addJavascriptInterface} but is unable to locate the corresponding URL/JavaScript-invocation site, e.g., a call to the \texttt{loadURL} method. Dynamic analysis successfully resolved these entries for 115 apps.

To illustrate the difference between the two phases, Table~\ref{tab:sample-entries} shows representative entries from the \textsc{QuickNote} app. The static phase resolves simple constant targets (row~1), but fails when the target depends on runtime state (row~2). The dynamic phase captures the actual values observed at runtime, including the concrete arguments that the static phase could not resolve (row~3).

\begin{table}[tbp]
	\centering
	\caption{Impact of guided dynamic analysis.}
	\label{tab:dynamic-summary}
	\begin{tabular}{lr}
		\hline
		\textbf{Metric} & \textbf{Value} \\
		\hline
		Runtime calls observed / retained & 668 / 662 \\
		Apps with $\geq$1 runtime witness & 211 / 309 (68.3\%) \\
		String-resolution entries (static / dynamic) & 444 / 662 \\
		Final entries (full corpus) & 1,106 \\
		Resolved at high confidence & 824 (74.5\%) \\
		New insertions / in-place upgrades & 577 / 85 \\
		Unresolved entries before / after dynamic & 248 / 207 (--16.5\%) \\
		Apps bootstrapped from empty & 115 / 155 \\
		\hline
	\end{tabular}
\end{table}

\begin{findingbox}
\textbf{Answer to RQ2.}
The dynamic phase adds 662 entries to the 444 produced by static analysis, raising the overall resolution rate to 74.5\%. It contributes all observed \texttt{evaluateJavascript} targets and provides first-time evidence for 115 apps that static analysis could not detect.
\end{findingbox}

\subsection{RQ3: How does \approach{} compare with IWandroid?}
\label{sec:rq3}

We compare \approach{} against \textsf{IwanDroid},~\cite{ifc-analysis-issre}, the closest related work. \textsf{IwanDroid} is a purely static, demand-driven analysis for the information flow analysis via the WebView bridge. To keep the comparison fair, we ran it on our 338 apps. \textsf{IwanDroid} could successfully analyze only 126 of the 338 apps. For string resolution, \textsf{IwanDroid} covers 112 packages versus 303 for \approach.

On the 126 apps where both tools produce output, \approach{} detects 455 bridge-registration records compared with 280 for \textsf{IwanDroid} ($1.63\times$). At the bridge-initiating class level, 171 classes are detected by both tools, 46 are found only by \approach, and 10 are found only by \textsf{IwanDroid}. The \textsf{IwanDroid}-only classes are exclusively Cordova \texttt{InAppBrowser} variants that do not register \texttt{@JavascriptInterface} methods; they are detected by IWandroid as WebView-hosting classes but are not true bridge registrations.

\textsf{IwanDroid} produces 249 distinct resolution entries, of which only 83 (33.3\%) are successfully resolved (a \texttt{const-string} value).  The remaining 66.7\% are explicit failure messages: 88 ``\emph{We do not handle inter-procedural as of now}'' (35.3\%), 39 ``\emph{move-result-object need to handle it}'' (15.7\%), and 39 other partially extracted values (15.7\%). In contrast, \approach{} produces 1,106 entries with 824 resolved at high confidence (74.5\%). Even when restricted to its static phase alone, \approach{} resolves 162 of 444 entries precisely (36.5\%)---already comparable to \textsf{IwanDroid}'s best case---and the dynamic phase then raises the total resolved count to 824.

\textsf{IwanDroid} produces zero \texttt{evaluateJavascript}-related entries. \approach{} captures 246, all from the dynamic phase.
Since \texttt{evaluateJavascript} is one of the key host-to-guest communication channels, this gap means \textsf{IwanDroid} misses an entire class of cross-language interactions. Table~\ref{tab:comparison} summarizes the comparison. \approach{} covers $2.68\times$ as many apps, produces $4.44\times$ as many resolution entries, and resolves $9.93\times$ as many of them. The advantage comes from three sources: (i)~a more robust static front-end that handles inter-procedural string propagation, (ii)~guided dynamic refinement that resolves targets invisible to any static analysis, and (iii)~\texttt{evaluateJavascript} support, which IWandroid lacks entirely.

\begin{table}[tbp]
\centering
\caption{Head-to-head comparison of \approach{} and IWandroid on 338 bridge-bearing apps.}
\label{tab:comparison}
\begin{tabular}{lrr}
\hline
\textbf{Metric} & \textbf{\approach} & \textbf{\textsf{IwanDroid}} \\
\hline
Apps with bridge records & 338 & 126 \\
Bridge-registration records & 1,047 & 280 \\
Packages with resolution entries & 303 & 112 \\
Distinct resolution entries & 1,106 & 249 \\
Resolved entries & 824 (74.5\%) & 83 (33.3\%) \\
Unresolved entries & 176 (15.9\%) & 166 (66.7\%) \\
\texttt{evaluateJavascript} entries & 246 & 0 \\
\hline
\end{tabular}
\end{table}

\begin{findingbox}
\textbf{Answer to RQ3.}
On the same dataset, \approach{} substantially outperforms \textsf{IwanDroid} in both coverage and resolution quality: it analyzes all 338 bridge-bearing apps (vs.\ 126), resolves 74.5\% of entries at high confidence (vs.\ 33.3\%), and is the only tool that captures \texttt{evaluateJavascript} targets.
\end{findingbox}
\subsection{RQ4: How effective is \approach at detecting inter-language bugs?}
\label{sec:rq4}

We evaluated \textsf{MifDroid}'s ability to expose inter-language flows to three JavaScript-analysis backends: CodeQL, Semgrep, and WebViewJSDetect. \Cref{tab:tools} identifies each backend and its tool category. Because CodeQL and Semgrep do not natively recognize \textsf{MifDroid}'s instrumentation markers, we configured their analyses to treat these markers as sources and sinks. The aggregate results are reported in \Cref{tab:summary}, and \Cref{tab:benchmark-results} provides the corresponding per-application breakdown.

\Cref{tab:summary} reports, for each backend, the number of applications with findings, the true- and false-positives values, and the precision score. To compute the precision score, we manually inspected the reported flows to assign these classifications. Both authors independently evaluated the reported vulnerabilities.

CodeQL produced 30 unique vulnerabilities, out of which 20 were classified as TPs and 20 as FPs, yielding $66.67\%$ precision. Semgrep produced 369 findings across 13 applications, all classified as FPs because its rules matched instrumentation markers rather than confirmed vulnerabilities. WebViewJSDetect produced no findings, for which precision is not defined. The per-application results in \Cref{tab:benchmark-results} show how these totals are distributed across the evaluated applications and report each backend's total findings, TPs, and FPs. Together, the tables show that CodeQL identifies confirmed flows in a subset of applications, whereas the Semgrep findings require further refinement to distinguish marker matches from vulnerabilities; WebViewJSDetect reports no flows in this benchmark.

\begin{table}[t]
	\centering
	\caption{Tools evaluated.}
	\label{tab:tools}
	\begin{tabular}{ll}
		\toprule
		\textbf{Tool}   & \textbf{Type}      \\
		\midrule
		CodeQL          & Commercial         \\
		Semgrep         & Commercial         \\
		WebViewJSDetect & Research prototype \\
		\bottomrule
	\end{tabular}
\end{table}

\begin{table}[t]
    \centering
    \caption{Summary statistics of detection results.}
    \label{tab:summary}
    \begin{tabular}{l|cc|ccc|c}
        \toprule
        \textbf{Tool}         &
        \textbf{Apps flagged} & \textbf{Apps w/ TP} &
        \textbf{\#}           & \textbf{TP}         & \textbf{FP} &
        \textbf{Precision (\%)}                                                            \\
        \midrule
        CodeQL                & 14                  & 11          & 30  & 20 & 10  & 66.67 \\
        Semgrep               & 13                  & 0           & 369 & 0  & 369 & 0.00  \\
        WebViewJSDetect       & 0                   & 0           & 0   & 0  & 0   & --    \\
        \bottomrule
    \end{tabular}
\end{table}

\begin{table}[t]
    \centering
    \caption{Detection results on benchmark applications.}
    \label{tab:benchmark-results}
    \begin{tabular}{l|ccc|ccc|ccc}
        \toprule
        \textbf{Application}                       &
        \multicolumn{3}{c|}{\textbf{CodeQL}}       &
        \multicolumn{3}{c|}{\textbf{Semgrep}}      &
        \multicolumn{3}{c}{\textbf{WebViewJSDetect}}                                                                   \\
        \cmidrule(lr){2-4}
        \cmidrule(lr){5-7}
        \cmidrule(lr){8-10}
                                                   & \textbf{\#} & \textbf{TP} & \textbf{FP}
                                                   & \textbf{\#} & \textbf{TP} & \textbf{FP}
                                                   & \textbf{\#} & \textbf{TP} & \textbf{FP}                           \\
        \midrule
        at.tomtasche.reader                        & 1           & 1           & 0           & 0  & 0 & 0  & 0 & 0 & 0 \\
        au.id.micolous.farebot                    & 1           & 1           & 0           & 4  & 0 & 4  & 0 & 0 & 0 \\
        be.lionslink.ucllstudent                  & 2           & 2           & 0           & 1  & 0 & 1  & 0 & 0 & 0 \\
        ch.gassenarbeit.bern.your.rights          & 2           & 2           & 0           & 1  & 0 & 1  & 0 & 0 & 0 \\
        com.a10miaomiao.bilimiao                  & 1           & 1           & 0           & 45 & 0 & 45 & 0 & 0 & 0 \\
        com.activitymanager                       & 1           & 0           & 1           & 1  & 0 & 1  & 0 & 0 & 0 \\
        com.app.openlib                            & 1           & 1           & 0           & 48 & 0 & 48 & 0 & 0 & 0 \\
        com.concealnetwork.concealmobile           & 5           & 1           & 4           & 1  & 0 & 1  & 0 & 0 & 0 \\
        com.cradle.iitc\_mobile                    & 6           & 5           & 1           & 4  & 0 & 4  & 0 & 0 & 0 \\
        com.duckduckgo.mobile.android              & 2           & 2           & 0           & 73 & 0 & 73 & 0 & 0 & 0 \\
        com.eightsines.firestrike.opensource       & 3           & 3           & 0           & 1  & 0 & 1  & 0 & 0 & 0 \\
        com.github.samotari.paynoway               & 2           & 0           & 2           & 58 & 0 & 58 & 0 & 0 & 0 \\
        com.in.my.district                        & 2           & 0           & 2           & 66 & 0 & 66 & 0 & 0 & 0 \\
        com.komodoplatform.komodowallet            & 1           & 1           & 0           & 66 & 0 & 66 & 0 & 0 & 0 \\
        \bottomrule
    \end{tabular}
\end{table}

\paragraph{Case Study: Bridge-to-Sink Flows in IITC Mobile}
\label{sec:case-study-bridge}
\begingroup\sloppy

\approach discovered a bridge-to-sink data flow with the CodeQL backend in the IITC mobile app (\texttt{com.cradle.iitc\_mobile}, version code 104), a client for the Ingress Intel map. The app binds an \texttt{IITC\_JSInterface} object to the name \texttt{android}. CodeQL found the two flows once MifDroid's instrumentation markers were modeled as sources, so that every bridge return value counts as untrusted. In both cases MifDroid's type check labels the Java method \emph{leaksData} but not \emph{carriesConfidentialValues}. Both flows are real data flows across the bridge, but neither is exploitable as reported. The interesting part is why, and what the second flow reveals once the Java side is examined. To assess the effectiveness of the analysis, we ran the application with native JavaScript-aware CodeQL, but it still failed to detect the bug.

\begin{lstlisting}[breaklines=true,basicstyle=\ttfamily\footnotesize,frame=single,numbers=left,label={lstlisting:foundscript}]
if(typeof android !== 'undefined' && android && android.getFileRequestUrlPrefix) {
  window.requestFile = function(callback) {
    do {
      var funcName = "onFileSelected" + parseInt(Math.random()*0xFFFF).toString(16);
    } while(window[funcName] !== undefined)
    window[funcName] = function(filename, content) {
      callback(decodeURIComponent(filename), atob(content));
    };
    var script = document.createElement('script');
    script.src = android.getFileRequestUrlPrefix() + funcName;
    (document.body || document.head || document.documentElement).appendChild(script);
  };
}
\end{lstlisting}

IITC lets plugins ask the user for a file. The JavaScript side builds a one-shot global
callback and loads a script from a URL whose prefix comes from the bridge. A code snippet from this is shown in~\cref{lstlisting:foundscript}. The instrumented stub returns \texttt{\_\_leaked\_\_(result)}, so CodeQL tracks the return value of \texttt{getFileRequestUrlPrefix()} into \texttt{script.src}. It reports the flow as both \texttt{js/xss} and \texttt{js/client-side-unvalidated-url-redirection}. Taken at face value, whoever controls this value controls which code runs in the WebView. On the Java side, the method delegates to \texttt{IITC\_FileManager.getFileRequestPrefix()}, which returns the constant \texttt{"//file-request.iitcm.localhost/"}. The forward flow therefore carries no attacker-influenced data, and the alert is not exploitable as stated. Following the request further shows a different flow. The URL is protocol-relative and points at a host the app serves itself. \texttt{IITC\_WebViewClient.shouldInterceptRequest} hands every \texttt{http(s)} request to a \texttt{*.iitcm.localhost} host to the file manager. For the \texttt{file-request} host, a \texttt{FileRequest} stores the first URL path segment as \texttt{mFunctionName} and opens a system file chooser. After the user picks a file, it streams back a script of the form

\begin{lstlisting}[breaklines=true,basicstyle=\ttfamily\footnotesize,frame=single]
<mFunctionName>('<URL-encoded filename>', '<Base64 file content>');
\end{lstlisting}

The filename is URL-encoded and the content Base64-encoded, but \texttt{mFunctionName} is
inserted verbatim. \texttt{Uri.getPathSegments()} also percent-decodes it. The callback name thus makes a round trip, i.e, from JavaScript to Java $\rightarrow$ executed JavaScript, with no validation. This is the callback-injection pattern familiar from JSONP. In the main frame it grants nothing new, because a script that can choose the request path can already execute code. The Java side does not check which frame issued the request, so any content that can make the WebView fetch this host can trigger the file chooser and supply the callback. It still cannot read a file unless the user selects one. We therefore rate the round-trip weakness low. It matters methodologically more than practically, because it is a JS-to-Java-to-JS flow, and none of the analyses in this paper models that direction.

\begin{findingbox}
\textbf{Answer to RQ4.}
\approach with CodeQL backend is able to find 30 vulnerabilities with the instrumented JavaScript, with a precision of $66.67\%$, while, Semgrep was ineffective in finding the vulnerabilities due to the limitations of the tool. This shows that \approach is effective in leveraging the strength of JavaScript only analysis to analyse information flow vulnerabilities across Android WebView applications. Four of the detected vulnerabilities were reported and confirmed by the developers.
\end{findingbox}

\subsection{Limitations and Threats to Validity}
\label{sec:limitations}
The coverage of the dynamic analysis phase is one threat to internal validity. In 98 cases, the dynamic analysis produced no meaningful results, i.e.,  the relevant WebView state is not reached within the bounded exploration budget. Thus, the reported gains should be interpreted as a lower bound of what hybrid design could produce.

A further limitation concerns host data embedded in injected scripts. \approach{} recovers the code of scripts injected through \texttt{evaluateJavascript} and \texttt{loadUrl("javascript:...")} and passes it to the JavaScript analyzer, but it does not track host values spliced into the script string, as in \texttt{evaluateJavascript("handle('" + token + "')")}. The recovered script contains the value of \texttt{token} only as a string constant, and the type-and-effect system treats the injection as $\mathbf{skip}$ (\textsc{T-Eval}), so no source marker is attached to it. Confidential host data that reaches the guest through this channel, rather than through the return value of a bridge method, is therefore not reported.

Another threat is how well our results generalize. We evaluate \approach only on the open-source F-Droid dataset, which does not necessarily reflect commercial apps, where obfuscation, proprietary SDKs, and deployment practices can be quite different. We mitigate this threat in part by evaluating on a large and diverse set of real F-Droid apps. In addition, the core challenges addressed by our approach, such as dynamic bridge targets and Java--JavaScript communication, are not unique to open-source apps.

\section{Related Work}\label{sec:related}

Yuan~\etal~\cite{webViewJsDetect} proposed \textsf{WebViewJSdetect}, which uses concurrent abstract interpretation to model the dynamic features of JavaScript within WebViews. Their approach navigates the dynamic nature of JavaScript analysis and is complementary to our work. Tiwari \etal~\cite{ifc-analysis-issre} proposed a demand-driven information flow analysis for WebView bridges in Android hybrid apps with limited string resolution. Their approach relies on a pre-existing call graph that does not model WebView-specific operations, which can lead to imprecision. In contrast, our approach performs more precise string resolution and constructs the call graph using dynamic analysis. Additionally, our modular design translates inferred effects into JavaScript stubs that can be consumed by existing single-language JavaScript security analyzers.
Hu \etal~\cite{omegatest} introduced $\omega$Test, a WebView-oriented testing technique that uses dynamic event-sequence generation to detect WebView bugs. However, $\omega$Test targets bug detection rather than information flow analysis. Our approach instead uses static analysis to guide dynamic analysis and construct per-WebView summaries for inter-language information flow analysis. Polycruise~\cite{polycruise} is a dynamic analysis framework for cross-language interactions, but it does not focus on information flow analysis or the specific semantics of WebView bridges. ReUnify~\cite{reunify} performs cross-language static analysis for JavaScript and Java in the context of react native apps.
Lee \etal~\cite{hybdridroid} proposed \textsf{HybriDroid}, which constructs an interprocedural CFG for hybrid apps but provides only limited support for IFC analysis, as its primary focus is detecting type-mismatch vulnerabilities. Tiwari \etal also proposed \textsf{LuDroid}~\cite{ludroidconference} for IFC analysis of hybrid apps, but it relies on coarse assumptions (e.g., treating any shared bridge function as leaking), which significantly reduces precision. In contrast, our approach performs explicit dataflow tracking from sources to sinks without such conservative assumptions.
\textsf{AdLib}~\cite{adlib2019ryu}, which analyzes data leaks in advertising libraries, is complementary to our work. Our framework can incorporate summaries of ad libraries augmented with integrity and confidentiality labels to generate the stubs for analysis.
Bae \etal~\cite{hydriddroidicse2019} formalized aspects of WebView semantics but focused on type-related vulnerabilities in bridge interfaces rather than information flow.
Prior work has explored WebView security from multiple angles, including code-injection and cross-zone vulnerabilities~\cite{rizzo2017babelview,chin2013vunerabilities}, application-specific analysis~\cite{mandal2018vulnerability}, and origin-related flaws~\cite{yang2018study}. Datta~\etal~\cite{datta_ndss_2026_webview} studied cross-boundary tracking via Java-to-JavaScript information diffusion, while Zhang~\etal~\cite{identity-confusion-webview} focused on identity confusion vulnerabilities. In contrast, our work provides a general IFC-based framework to detect information flow violations across the bridge. Beyond WebViews, Buro~\etal~\cite{buro2020abstractmultlingual,buro2023multilanguage} develop a general theory of multi-language semantics and abstraction; \approach{} instantiates a bridge-centric decomposition for the concrete case of Android WebViews, where the bridge is configured dynamically. Closer to our modular design, Prakash~\etal~\cite{confluence-scp} unify unilingual pointer analyses for multilingual FFI-based programs, and Roth~\etal~\cite{roth2024axa} propose \textsf{AXA}, which obtains a cross-language analysis by integrating existing single-language analyses. Like these approaches, \approach{} reuses single-language analyzers rather than building a joint one; it differs in exchanging information-flow summaries across the WebView bridge as JavaScript stubs.

\section{Conclusion}\label{sec:conclusion}
We presented \approach{}, a hybrid analysis framework for detecting inter-language information flow violations in Android hybrid apps. We demonstrate that a combination of static and dynamic analysis achieves good coverage in resolving WebView targets. Modular design of \approach{} helped in detecting 14 vulnerable cross-language information flows and 45 arity mismatches, demonstrating that the hybrid design yields better results than either phase alone.

\begin{findingbox}
 \textbf{Data Availability}
The prototype implementation of \approach and the detailed experimental results are available at: 
\url{https://doi.org/10.5281/zenodo.19248381}. 

\end{findingbox}

\bibliographystyle{ACM-Reference-Format}
\bibliography{references}

\end{document}